\documentclass[11pt]{JHEP3}

\pdfoutput=1

\usepackage{slashed}

\usepackage{color}

\usepackage{epsfig}
\usepackage{epstopdf}
\usepackage{epsf}
\input{epsf.sty}
\usepackage{graphicx,amsmath,amssymb}
\usepackage{epsfig,multicol}
\allowdisplaybreaks

\usepackage{bbm,bm,amsmath,amssymb}

\def\bg{\begin{eqnarray}}
\def\nd{\end{eqnarray}}
\def\sin{{\rm sin}}
\def\cos{{\rm cos}}

\def\log{{\rm log}}

\newcommand{\be}{\begin{equation}}
\newcommand{\ee}{\end{equation}}
\newcommand{\ba}{\begin{eqnarray}}
\newcommand{\ea}{\end{eqnarray}}

\newcommand{\lp}{\left(}
\newcommand{\rp}{\right)}
\newcommand{\ls}{\left[}
\newcommand{\rs}{\right]}

\def\rmi{{\rm i}}

\def\ib{{\bar \imath}}
\def\jb{{\bar \jmath}}

\newcommand{\zp}{0}
\newcommand{\zm}{{\bar 0}}
\newcommand{\ip}{i}
\newcommand{\jp}{j}
\newcommand{\im}{\ib}
\newcommand{\jm}{\jb}

\newcommand{\ft}[2]{{\textstyle\frac{#1}{#2}}}

\title{Fermions on the Anti-Brane: Higher Order Interactions and Spontaneously Broken Supersymmetry} 

\author{Keshav Dasgupta, Maxim Emelin, and Evan McDonough\\
\vskip.03in
Ernest Rutherford Physics Building, McGill University,\\
3600 University Street, Montr{\'e}al QC, Canada H3A 2T8\\
{\tt keshav@hep.physics.mcgill.ca, maxim.emelin@mail.mcgill.ca}\\
{\tt evanmc@physics.mcgill.ca}}

\date{\today}

\abstract{It has been recently argued that inserting a probe ${\overline{\rm D3}}$-brane in a flux background breaks supersymmetry spontaneously instead of explicitly, as previously thought. In this paper we argue that such spontaneous breaking of supersymmetry persists even when the probe ${\overline{\rm D3}}$-brane is kept in a curved background with an internal space that doesn't have to be a Calabi-Yau manifold. To show this we take a specific curved background generated by fractional three-branes and fluxes on a non-K\"ahler resolved conifold where supersymmetry breaking appears directly from certain world-volume fermions becoming massive. In fact this turns out to be a generic property even if we change the dimensionality of the anti-brane, or allow higher order fermionic interactions on the anti-brane. We argue for the former by taking a probe ${\overline{\rm D7}}$-brane in a flux background and demonstrate the spontaneous breaking of supersymmetry using world-volume fermions. We argue for the latter by constructing an all order fermionic action for the ${\overline{\rm D3}}$-brane from which the spontaneous nature of supersymmetry breaking can be demonstrated by bringing it to a $\kappa$-symmetric form.}

\maketitle

\begin{document}

\section{Introduction}

It has recently been shown \cite{wrase1, wrase2}
that a probe $\overline{\rm D3}$-brane in a flux background breaks supersymmetry spontaneously, and furthermore, if the $\overline{\rm D3}$ is placed on an orientifold plane, the only low-energy field content is a single massless 
fermion\footnote{See also \cite{shiu, otherrefs}, and especially the key papers \cite{VolkovA}, that motivated the research on spontaneous susy breaking in the presence of a 
${\overline{\rm D3}}$-brane.}. 
The implications of this are two-fold: (1) that SUSY breaking is spontaneous, as opposed to explicit, indicates that there is no perturbative instability in the D3-$\overline{\rm D3}$ system famously used to construct the KKLT de Sitter solution \cite{Kachru:2003aw}, and (2) as the only 
four-dimensional field content is a single massless fermion, which can be expressed in the $d=4$ $\mathcal{N}=1$ supergravity theory as the spinor component of a nilpotent multiplet, this provides a natural starting point for a string theory embedding of the inflation models proposed in \cite{Kallosh:2014via,Ferrara:2014kva, McDonough:2016der} and other works.

This result, and the connection to string cosmology, provides impetus to further investigate $\overline{\rm Dp}$-brane systems; in order to populate the landscape of stable non-supersymmetric compactifications with $\overline{\rm Dp}$-branes, to better understand supersymmetry breaking in these models, and to perhaps stumble upon new string theory settings where de Sitter space and inflation naturally arise. It is with these goals in mind that we present three interconnected analyses, which generalize and build upon the work of \cite{shiu, wrase1, wrase2}.

\subsection{Spontaneous vs. explicit supersymmetry breaking with anti-branes}
\label{sec:spont}

Before we proceed with our analysis, let us start with a discussion of spontaneous supersymmetry breaking.

Spontaneous supersymmetry breaking is a crucial element of string theory model building. This is because a consistent study of four dimensional physics requires that all or almost all moduli be stabilized, and all known mechanisms of moduli stabilization\footnote{with the exception of `string gas' moduli stabilization, see e.g. \cite{Brandenberger:2005fb}} are understood in terms of a supersymmetric four dimensional theory, e.g. the complex structure moduli are fixed via the flux induxed superpotential as in \cite{GKP}. Without an underlying supersymmetric theory, i.e. in the case that supersymmetry is explicitly broken, it is not clear to what extent the known methods of moduli stabilization are applicable.

Spontaneous symmetry breaking occurs when the ground state of a theory does not respect the symmetries of the action. This is an essential part of model building in particle physics, supergravity, and string theory, as it gives theoretical control over corrections to the action. The situation in string theory is slightly more complicated than in particle physics, since proposed de Sitter solutions in string theory (for example KKLT \cite{kklt}) rarely exist as the ground state of the theory, but rather as metastable minima. Given this, we will drop the phrase `ground state' from our definition, and instead refer to non-supersymmetric states in a supersymmetric theory as spontaneously breaking the supersymmetry.

In simple cases, for example \cite{wrase2}, there is a smoking gun of spontaneous supersymmetry breaking by antibranes: a worldvolume fermion remains massless, which one can identify with the goldstino of SUSY breaking. However, as discussed in \cite{shiu}, it will not in general be true that a worldvolume fermion remains massless. Instead, the goldstino of SUSY breaking will be some combination of open and closed string modes. Thus a more general diagnostic of spontaneous breaking is needed, which we will now develop. We will see that even in the absence of a massless fermion on the brane, supersymmetry breaking can still be shown to be spontaneous.

Our diagnostic for spontaneous supersymmetry breaking by a probe $\overline{\rm Dp}$ brane is the following: a solution breaks supersymmetry spontaneously if it is a solution of the theory with action:
\begin{equation}
\label{susyaction}
S = S_{\rm IIB} + S_{\overline{\rm Dp}} ,
\end{equation}
where $S_{\rm IIB}$ is action of type IIB supergravity. The above action is explicitly supersymmetric, since an anti-brane is 1/2 BPS, and thus negates the requirement to `find' the goldstino in order to deduce that supersymmetry breaking is spontaneous. A probe $\overline{\rm D3}$ in a non-compact GKP background without sources can be studied in this way.  This reasoning applies directly to our second example:  an $\overline{\rm D7}$ in a warped bosonic background without sources, which we will study in Section 3.

However, this diagnostic is limited in its applicability, as many interesting backgrounds have explicit brane or orientifold content in addition to the probe $\overline{\rm Dp}$. Fortunately, the condition (\ref{susyaction}) can in fact be extended to apply to a subset of these cases, by making use of string dualities to relate a flux background with branes to a background without branes. Again, this makes no recourse to the goldstino being a pure open-string mode, i.e. a worldvolume fermion.

Our first example in this paper, a $\overline{\rm D3}$ in a resolved conifold background with wrapped five-branes, which we study in Section 2, is an example where dualities must be used to make sense of \eqref{susyaction}. One way to arrive at the resolved conifold with wrapped five-branes background is as a solution to $S=S_{\rm IIB}+ S_{\rm D5}$, in which case the addition of a $\overline{\rm D3}$ would break supersymmetry explicitly, since the D5 and $\overline{\rm D3}$ are invariant under different $\kappa$-symmetries. However, the resolved conifold background can alternatively be found as the dual to the deformed conifold with fluxes and no branes\footnote{The dual is succinctly described in supergravity when the number of 
wrapped D5-branes is very large \cite{vafa, becker}.}, 
see for example \cite{becker, Gwyn:2007qf}. In this dual frame the underlying action is source-free, and the addition of an $\overline{\rm D3}$ (again in the dual deformed conifold) will break SUSY spontaneously. The deformed conifold with $\overline{\rm D3}$ can then dualized back to a resolved conifold with wrapped ${\rm D5}$ along with a $\overline{\rm D3}$, but the spontaneous (as opposed to explicit) nature of SUSY breaking is only manifest in the dual frame.

As we will see, backreaction of the $\overline{\rm D3}$ on the resolved conifold induces masses for all the fermions, so there is no obvious candidate for the goldstino; this further indicates that the resolved conifold with wrapped D5 and a $\overline{\rm D3}$ system exhibits explicit breaking of supersymmetry. This is consistent with our discussion above: the spontaneous nature of SUSY breaking is only manifest in the dual deformed conifold description. In terms of moduli stabilization, a dual description in terms of spontaneous breaking allows one to consistently define a superpotential for both the K\"{a}hler and complex structure moduli, which is precisely the feature of `spontaneous breaking' that is useful for studying $4d$ physics from string theory.

\subsection{Outline of the paper}

Our first analysis, studied in section 2, considers a probe $\overline{\rm D3}$-brane, not in a Calabi-Yau background \cite{DRS, GKP}  as studied in \cite{wrase2}, but in a non-K\"ahler resolved conifold background with integer and fractional three-branes. We will construct a supersymmetric deformation to the Calabi-Yau resolved conifold that converts it to a non-K\"ahler resolved conifold, provides a non-zero curvature to the internal space, and which induces a non-zero amount of ISD fluxes. Once a probe $\overline{\rm D3}$ is introduced, supersymmetry is spontaneously broken by the coupling of ISD fluxes to the worldvolume fermions, giving masses to the world-volume fermions. This breaking is in fact `soft' as the fluxes and fermion masses are set by the non-Kahlerity of the internal space, which is in turn a tune-able parameter. The picture is somewhat similar to the case with Calabi-Yau internal space as studied in \cite{wrase2} but the analysis differs in terms of fluxes and backreaction. In particular, the analysis in the probe approximation now yields \emph{two} massless fermions, as opposed to one in \cite{wrase2}. This result is modified upon considering backreaction of the $\overline{\rm D3}$ on the bulk fluxes, which generates both (2, 1) and (1, 2) three-form fluxes, inducing masses for \emph{all} the worldvolume fermions, i.e. there are \emph{zero} massless fermions remaining in the spectrum. We also study certain aspects of de Sitter vacua from our analysis.  It interesting to note that a curved internal space appears to be a requirement for de Sitter solutions in string theory, at least in many contexts,  especially negatively curved internal spaces (see for example \cite{Danielsson:2011au} and references therein). With this in mind, we consider moduli stabilization in this background, and the connection to de Sitter space in this model.

The physics discussed above remains largely unchanged even if we change the dimensionality of the anti-brane. In section 3, we consider a second application of anti-brane fermionic actions and take a probe\footnote{By assuming such a heavy object as probe simply means that the logarithmic backreactions of the ${\overline{\rm D7}}$-brane on geometry and fluxes are suppressed by powers of $g_s$.}  $\overline{\rm D7}$-brane, this time working with a Calabi-Yau background. Supersymmetry is again broken spontaneously via flux-induced fermion masses, and the masses are proportional to the piece of the three-form flux which is ISD in the space transverse to the brane. In the $\overline{\rm D3}$ case, where the transverse space is the entire internal space, this flux is precisely the flux of the GKP background\footnote{Henceforth by GKP background we will always mean the background proposed in \cite{DRS, GKP}.}. 
However, in the $\overline{\rm D7}$ case, the fermion masses are now sourced by the subset of these fluxes which are ISD in the two-directions transverse to the brane. In other words, the fermion masses are now determined solely by fluxes that have two legs on the brane, and one leg off.  We show that for a special class of flux background there can be many massless fermions in the low energy spectrum, while in a general flux background there may be none. This provides yet another instance of a string theory realization of nilpotent goldstinos\footnote{See  \cite{Kallosh:2015nia, Bertolini:2015hua, Garcia-Etxebarria:2015lif} for even more examples.}, 
and a possible starting point for inflation and de Sitter solutions.

Our final application is actually  closer to a derivation; we study the fermionic $\overline{\rm D3}$ action at \emph{all} orders in the fermionic expansion. To do this, we promote the bosonic fields to superfields, and discuss the physics at the self-dual point. At the self-dual point we can use U-dualities to relate various pieces of the multiplet and consequently determine the fermionic completions of the different fields. Once we move away from the self-dual point, we can determine the fermionic completions of all the bosonic fields in a compact form. As an added bonus, we find that the all-order fermionic action can be written in a manifestly $\kappa$-symmetric form, even without precise details of the form of the terms in the action. The orientifolding action can then be easily incorporated in the action. This indicates that the spontaneous nature of supersymmetry breaking by anti-branes, both in the presence and in the absence of an orientifold plane, is not a leading order effect, but in fact continues to be true to all orders. This puts the conclusions of \cite{wrase1, wrase2}, and its implications for KKLT, on solid footing.

We conclude with a short discussion of the implications of our work and directions for future research.

\section{$\overline{\rm D3}$-brane in a Resolved Conifold Background: Soft (and Spontaneous) Breaking of Supersymmetry}

The breaking of supersymmetry by a probe $\overline{\rm D3}$-brane in a warped bosonic background was studied recently in \cite{wrase2}. They studied a ${\overline{\rm D3}}$-brane in a GKP background, and found that supersymmetry was spontaneously broken by the coupling of ISD fluxes to the worldvolume fermions. In this section we perform a similar analysis, focusing instead on a probe ${\overline{\rm D3}}$-brane in a  resolved conifold background. We will consider a deformation to the Calabi-Yau resolved conifold which maintains supersymmetry but provides a non-zero curvature to the internal space, and which induces ISD three-form fluxes from a set of integer and fractional D3-branes. Once a probe $\overline{\rm D3}$ is introduced, supersymmetry is again 
spontaneously (and softly) broken by the coupling of ISD fluxes to the worldvolume fermions, and the fermion masses can be straightforwardly computed. As we will see, the `soft' nature of supersymmetry breaking is due to the tune-able nature of the non-K\"{a}hlerity of the internal manifold.

The key details of the fermionic action for a ${\overline{\rm D3}}$-brane in a warped bosonic background are given in \cite{wrase2}. These will be the starting point of our analysis, so here we merely quote them. The worldvolume action is given, in a convenient $\kappa$-symmetry gauge, by
\be\label{eq:antiD3action2}
{\cal L}_f^{\overline {\rm D3}} = T_3 e^{4 A_0} \, \bar{\theta}^1 \ls 2 e^{-\phi} \Gamma^\mu \nabla_\mu -\frac{i}{12}\lp {\cal G}^{{\rm ISD}}_{mnp} - \bar{{\cal G}}^{{\rm ISD}}_{mnp} \rp \Gamma^{mnp}\rs \theta^1\,.
\ee
where $\theta^1$ is a 16-component\footnote{16 complex components, or 32 real components.} $10d$ Majorana-Weyl spinor\footnote{We have already fixed $\kappa$-symmetry.}, and we have defined the three from flux ${\cal G}_3$ as ${\cal G}_{(3)} = F_{(3)} - \tau H_{(3)}$. The 16-component spinor $\theta^1$ can be decomposed into four $4d$ Dirac spinors $\lambda^0$, $\lambda^i$ with $i=1,2,3$. On a Calabi-Yau manifold, the $\lambda^0$ is a singlet under the $SU(3)$ holonomy group of the internal Calabi-Yau manifold while the $\lambda^i$ transform as a triplet. 

We can now rewrite the $\overline {\rm D3}$ brane action \eqref{eq:antiD3action2} using the $4d$ decomposition of the $\theta^1$ spinor in the following way:
\begin{eqnarray}
\label{antiD3fermionaction}
{\cal L}_f^{\overline {\rm D3}} &=&2 T_3 e^{4 A_0-\phi} \, \left[\bar \lambda_-^\zm\gamma^\mu\nabla _\mu\lambda_+^\zp+
  \bar \lambda_-^{\jm}\gamma^\mu\nabla _\mu\lambda_+^{\ip}\delta_{\ip\jm}\right. \label{eq:antiD3action4d}\\
 &&\left. \phantom{2 T_3 }+\ft12m_0 \bar \lambda_+^\zp\lambda_+^\zp +\ft12\overline{m}_0 \bar \lambda_-^{\zm}\lambda_-^{\zm}
  +m_i\bar \lambda_+^{\zp}\lambda_+^{\ip}+\overline{m}_{\ib}\bar\lambda_-^{\zm} \lambda_-^{\im}+\ft12 m_{ij}\bar \lambda_+^{\ip}\lambda_+^{\jp}+\ft12\overline{m}_{\ib\jb}\bar \lambda_-^{\im}\lambda_-^{\jm}\right]\,,
\nonumber
\end{eqnarray}
where we use $\pm$ subscripts to denote $4d$ Dirac spinors that satisfy $\lambda_\pm = \frac12(1 \pm i  {\widetilde{\Gamma}}_{0123}) \lambda$, and the masses are defined as
\begin{eqnarray}
&& m_0  = \frac{\sqrt{2}}{12}\rmi e^\phi \bar \Omega^{uvw}\bar{\cal G}^{{\rm ISD}}_{ u v  w}\,,~~ \qquad \qquad \qquad\qquad \qquad  \qquad \qquad \qquad  \mbox{from }(0,3)\mbox{ flux,}\label{eq:msing}\\
&& m_i =  -\frac{\sqrt{2}}{4}e^\phi e_i^u\,\bar{\cal G}^{{\rm ISD}}_{uv\bar w} J^{v\bar w}\,,~~~~\qquad \qquad \qquad \qquad \mbox{from non-primitive }(1,2)\mbox{ flux,}\label{eq:mmix}\\
&&  m_{ij} =  \frac{\sqrt{2}}{8}\rmi e^\phi\left(e_i^{w}e_j^t+e_j^{w}e_i^t\right)\Omega_{uvw}g^{u\bar u}g^{v\bar v}\bar{\cal G}^{{\rm ISD}}_{t \bar u\bar v}\,,\qquad 
\quad\mbox{ from primitive }(2,1)\mbox{ flux,}
 \label{eq:mtrip}
\end{eqnarray}
where  $J$ and $\Omega$ are the K\"ahler form and holomorphic 3-form respectively. 

We are interested in a more general background, where the $SU(3)$ holonomy will be broken by a perturbation to the geometry. Compactifications on manifolds with $SU(3)$ structure but not $SU(3)$ 
holonomy have been studied in, for example, \cite{Benmachiche:2008ma} and \cite{Gurrieri:2007jg}. These are non-K\"ahler manifolds, which in general may or may not have an integrable complex structure,
and are classified by five torsion classes ${\cal W}_i$ \cite{chiossi, luest, louis}. The simplest case, where all five torsion classes vanish, is a Calabi-Yau manifold that supports no fluxes.   We are looking for the case with fluxes, so that we can make use of equations \eqref{eq:msing}, \eqref{eq:mtrip}, and \eqref{eq:mmix}, 
and therefore some of the torsion classes must be non-zero. 

Moreover, the non-K\"ahler manifold that we need has to be 
a complex manifold, otherwise the flux decomposition in terms of (2, 1), (1, 2) or (0, 3) forms would not make any sense. In addition, the manifold should to be non-compact, so as to avoid any tension with 
Gauss' law. The simplest internal manifold that satisfies our requirements is the resolved conifold with a non-K\"ahler metric which allows an integrable complex structure (and by definition doesn't have a conifold singularity).

The goal of this section will be to study the action \eqref{eq:antiD3action2} or \eqref{antiD3fermionaction}
in a resolved conifold with an arbitrary amount of D3 branes and delocalized five branes (see \cite{Gaillard:2008wt} and \cite{DEM} for more details on delocalized sources). More precisely, we will put a $\overline{\rm D3}$-brane in a supersymmetric background with metric given by:
\bg\label{iibmeet}
ds^2 = {1\over e^{2\phi/3} \sqrt{e^{2\phi/3} + \Delta}} ~ds^2_{0123} + e^{2\phi/3} \sqrt{e^{2\phi/3} + \Delta} ~ds^2_6,
\nd
where $e^\phi$ is related to type IIB dilaton $e^\phi_B$ as $\phi_B = -\phi$  and the factor $\Delta$ encodes the backreaction of the 3-branes. It is defined using a parameter $\beta$ as:
\bg \label{Delta} \Delta = {\rm sinh}^2\beta \left(e^{2\phi/3} - e^{-4\phi/3}\right). \nd
The other piece appearing in \eqref{iibmeet} is 
$ds^2_6$, which is the metric of the internal six-dimensional non-K\"ahler resolved conifold. This is expressed in terms of the coordinates ($r, \psi, \theta_i, \phi_i$) in the following way:  
\bg\label{resolve}
ds^2_6 = F_1~ dr^2 + F_2 (d\psi + {\rm cos}~\theta_1 d\phi_1 + {\rm cos}~\theta_2 d\phi_2)^2  + \sum_{i = 1}^2 F_{2+i}
(d\theta_i^2 + {\rm sin}^2\theta_i d\phi_i^2),
\nd
where the resolution parameter is proportional to $F_3 - F_4$. 

We will start by making an ansatze for the warp-factors $F_i(r)$ appearing in \eqref{resolve} which will allow us to see how to go from a 
Ricci-flat Calabi-Yau metric to a non-K\"ahler metric on a resolved conifold. A more generic class of solutions for the warp-factors exists and has been discussed in \cite{DEM}, but we will only consider a 
subset given by:   
\bg\label{churamont}
F_1 = {1\over F} + \delta F, ~~~~~ F_2 = \frac{r^2 F }{9}, ~~~~~F_3 = {r^2\over 6} + a_1^2(r), ~~~~~ F_4 = {r^2\over 6} + a_2^2(r), ~~~~~ \phi=\phi(r) , \nd
where $F$, $\delta F(r)$, $a_1(r)$, and $a_2(r)$, are functions of the radial coordinate only. From the above ansatze, it is easy to see where the Calabi-Yau case fits in. It is given by:
\bg\label{cymet}
F(r) \equiv F_{CY } = \left({r^2 + 9a^2 \over r^2 + 6 a^2}\right),~~~~~ \delta F(r) = 0, ~~~~~ a_1(r) = a, ~~~~~ a_2(r) = 0, ~~~~~ \phi=0. \nd
The Calabi-Yau case is fluxless (with the vanishing of the flux enforced by supersymmetry), and has a constant dilaton. Once we switch on fluxes, we can no longer assume that the other pieces of the 
warp-factors appearing in \eqref{churamont} vanish. 

As a cautionary tale, let us first consider whether we can perturb away from Calabi-Yau resolved conifold simply by allowing for a small perturbation in $F(r)$ and $\phi(r)$. We will see that this in fact does not lead to useful results, and thus we will need to be more careful in constructing our geometry. Nonetheless, it is useful for establishing an algorithm for constructing solutions.

Consider a small perturbation to \eqref{cymet} of the form:
\bg\label{gorjon}
F(r) = F_{CY} + \sigma f(r), ~~~~~~ \delta F(r) = 0, ~~~~~~ a_1(r) = a e^{-\phi}, ~~~~~~ a_2(r) = 0, \nd
where $\sigma$ is a dimensionless expansion parameter, that satisfies the the EOMs and takes the solution from the Calabi-Yau resolved conifold to the non-K\"ahler resolved conifold.  We can narrow down our perturbation scheme by allowing the dilaton field to behave in the following way:
\begin{equation}
\label{phisol}
\phi(r)=\log \left(\frac{1}{r^\sigma} \right),
\end{equation}
which would guarantee the existence of a small parameter $\sigma$ that, while preserving supersymmetry, would be responsible in taking us away from the Calabi-Yau case. In the limit $\sigma \to 0$, 
we go back to the fluxless Calabi-Yau case. This geometry is of course singular in the $r \rightarrow \infty$ limit, but we will assume for this discussion that the geometry is capped off at some sufficiently large $r$. In any case, this issue will not be important, as this perturbation fails for other reasons.

A way to construct such a background has already been discussed in \cite{DEM}, and therefore we will simply quote some of the steps. The best and probably the easiest way to analyze such a background
is by using the torsion classes. For us the relevant torsion classes are ${\cal W}_4$ and ${\cal W}_5$. They can be expressed in terms of the warp-factors $F_i(r)$ and the dilaton $\phi(r)$ in the 
following way:
\begin{eqnarray}
\label{torsion}
&& \mathcal{W}_4  =  {F_{3r} - \sqrt{F_1 F_2}\over 4 F_3}  + {F_{4r} -  \sqrt{F_1 F_2}\over 4 F_4} + \phi_r , \nonumber \\
&& \mbox{Re} \,{\cal W}_5 = \frac{F_{3r}}{12 F_3} + \frac{F_{4r}}{12 F_4}  + \frac{F_{2r} - 2\sqrt{F_1 F_2 }}{12 F_2} + {\phi_r\over 2}. \end{eqnarray}
The other torsion classes take specific values, with ${\cal W}_3$ determining the torsion. This solution is generated by following the duality chain described in \cite{DEM}, which generates both the RR and the NS three-forms ${\cal F}_3$ and ${\cal H}_3$ respectively. 

Our aim then is to use these torsion classes to determine the functional form for the warp-factors $F_i$ using the specific variation of the ansatze \eqref{churamont} i.e \eqref{gorjon} and 
\eqref{phisol}. The key relation, that allows us to find the connection between $F(r)$ and the dilaton $\phi(r)$, is the supersymmetry condition:
\begin{equation}\label{susycondition} 2 \mathcal{W}_4 +\mbox{Re} \,{\cal W}_5  = 0.\end{equation}
Plugging in the ansatze \eqref{gorjon} and \eqref{phisol} in \eqref{susycondition} will allow us to determine $f(r)$ completely in terms of the radial coordinate $r$ and the resolution parameter 
$a^2$. The functional form for $f(r)$ turns out to be a non-trivial function of $r$:
\bg\label{fra}
f(r) & = & {2\over (6a^2 + r^2)}\Bigg\{27 a^2 (6a^2 + r^2)\left[\sum_{i=1}^3 \Phi_i(r; a^2) + r^2 {\rm log}~r\right] \nonumber\\ 
 &-&  (9a^2 + r^2)(6a^2+r^2)\left[ 3{\rm log}\left({r^2\over 6a^2}+1\right) + 2 - 
{r^2 {\rm log}~r\over 6a^2 + r^2}\right]\Bigg\}, \nd
which is defined for $a^2 > 0$. For vanishing $a^2$ the functional form for $f(r)$ simplifies and has been studied earlier in \cite{lapan}. The other variables appearing in \eqref{fra} are 
defined in the following way:
\bg\label{jotadhor}
&&\Phi_1(r; a^2) ~ = ~ {}_2F_1^{(0, 0, 1, 0)}\left(-1, 2, 3, -{r^2\over 6a^2}\right) ,\nonumber\\
&& \Phi_2(r; a^2) ~ = ~ {}_2F_1^{(0, 1, 0, 0)}\left(-1, 2, 3, -{r^2\over 6a^2}\right) ,\nonumber\\
&& \Phi_3(r; a^2) ~ = ~ {}_2F_1^{(1, 0, 0, 0)}\left(-1, 2, 3, -{r^2\over 6a^2}\right), \nd  
where the notation ${}_2 F_1^{(0,1,0,0)}$ refers to $ \partial_y \,{}_2 F_1[x;y;z;w]$, and similarly for  ${}_2 F_1^{(1,0,0,0)}$ and  ${}_2 F_1^{(0,0,1,0)}$.  
This perturbation to $F(r)$ corresponds to introducing a small Ricci scalar on the internal space. This could computed using the torsion classes (\cite{torsionSU3ricci}), or computed directly using standard GR techniques. Using GR techniques, we find a simple expression emerges for small resolution parameter $a^2$ and small value for the parameter $\sigma$:
\begin{equation}
\delta R_6 =  - \frac{72 \sigma}{r^2} \left[3 -  2 \log\left( \frac{6 a^2}{r^2}\right) \right],
\end{equation}
which is negative for $r \geq 1.2 a $. Furthermore one can check that for \emph{general} $a$, i.e. not small $a$, while the expression for $\delta R_6$ is no longer simple, it is negative definite.  
It is interesting to note that negatively curved internal spaces have been widely studied as a mechanism for finding de Sitter solutions in string theory, see the discussion and references in \cite{Danielsson:2011au}.

The above analysis, although interesting because of the control we can have on the non-K\"ahlerity of the internal manifold, is ultimately {\it not} useful for finding the masses of the ${\overline{\rm D3}}$ world-volume 
fermions, as it in fact renders the internal manifold with a non-integrable complex structure. Thus, there exists an almost complex structure but the manifold itself may 
not be complex\footnote{There might exist a non-trivial {\it integrable} complex structure, but we 
haven't been able to find one.}. This means we cannot decompose our $\mathcal{G}_3$ flux in terms of (1, 2), (2, 1) or (0, 3) forms in a global sense, making the fermionic mass decompositions
given in \eqref{eq:mtrip}, \eqref{eq:mmix} and \eqref{eq:msing}, 
not very practical in analyzing the fermions on the probe ${\overline{\rm D3}}$. This of course doesn't mean that we cannot study the spontaneous susy breaking; we can,
but the analysis will not be so straightforward as was with the complex decomposition of the three-form fluxes.

The question then is: can we have a {\it complex} non-K\"ahler resolved conifold satisfying a more generic ansatze like \eqref{churamont} where we can use equations
\eqref{eq:msing}, \eqref{eq:mtrip}, and \eqref{eq:mmix}, to study spontaneous susy breaking with a probe ${\overline{\rm D3}}$? The answer turns out to be in the affirmative, and in the 
following section we elaborate the 
story\footnote{Note that there is some subtlety with the mapping to \cite{pandoz} at this stage, for example the possibility of a non-K\"ahler special Hermitian solution with a constant dilaton that we get here demanding supersymmetry as opposed to a Calabi-Yau resolved conifold with a constant dilaton studied in \cite{pandoz}. This has been discussed in details in \cite{DEM} so we will not dwell on this any further.}.

\subsection{A SUSY perturbation of the resolved conifold}

Let us start with a simple example of a D3-brane located at a point in an internal manifold specified by the metric $ds^2_6$ where $ds^2_6$ is given by:
\bg\label{3met}
ds^2_6 = dr^2 + g_{mn} dy^m dy^n, \nd
where ($r, y^m$) are the coordinates of the internal six-dimensional space. To avoid contradiction with Gauss' law, the internal manifold has to be non-compact, although
a compact example could be constructed by either inserting orientifold planes, or anti-branes. Details of this will be discussed later. The backreaction of the D3-brane
converts the vacuum manifold:
\bg\label{vacuum}
ds^2_{\rm vac} = ds^2_{0123} + ds^2_6, \nd
with $ds^2_{0123}$ being the Minkowski metric along the space-time directions, to the following:
\bg\label{back}
ds^2_{10} = {1\over \sqrt{h}} ds^2_{0123} + \sqrt{h} ds^2_6, \nd
where $h$ is the warp-factor. The five-form flux in the background \eqref{back} is now given as:
\bg\label{5form}
{\cal F}_5 = {1\over g_s} \left(1+ \ast_{10}\right) dh^{-1} \wedge dx^4. \nd
The above analysis is generic, but it is highly non-trivial to actually compute the warp-factor $h$. For a complicated internal space, the equation for $h$ 
typically becomes an involved second-order PDE. Furthermore, in the presence of other type IIB fluxes, for example the three-form fluxes ${\cal H}_3$ and ${\cal F}_3$,
the metric is more complicated than \eqref{back}. Additionally, the string coupling constant generically will not be 
constant.

There is, however, a way out of the above conundrum if we analyze the picture from a more general setting. We can use the powerful machinery of torsional analysis 
\cite{luest, louis, gauntlett} to write the background of a D5-brane wrapped on some two-cycle, parametrized by ($\theta_1, \phi_1$),  
of a generic six-dimensional internal space. Assuming that the 
size of the wrapped cycle is smaller than some chosen 
scale, any fluctuations along the ($\theta_1, \phi_1$) will take very high energy to excite. This means at low 
energies the theory will be of an effective D3-brane\footnote{Also known as a fractional D3-brane. There is yet another way to generate a fractional D3-brane which we don't explore here. For 
example if we take wrapped D5-${\overline{\rm D5}}$-branes with ($n_1, n_2$) amount of gauge fluxes on each of them, then we can have bound D3-branes with charges $n_1$ and $n_2$ respectively. If
$n_i$ are fractional, these give fractional three-branes with vanishing global five-brane charges. See \cite{DSW, DM2} for more details.} 
and the source charge of the wrapped D5-brane $C_6$ will decompose as:
\bg\label{sourcech}
C_6(\overrightarrow{\bf x}, \theta_1, \phi_1) = C_4(\overrightarrow{\bf x}) \wedge\left({e_{\theta_1} \wedge e_{\phi_1}\over \sqrt{V}}\right), \nd
where $V$ is the volume of the two-cycle on which we have the wrapped D5-brane. Therefore using the criteria \eqref{sourcech}, the 
supergravity background for the configuration of the effective D3-brane is 
given by:
\bg\label{sugra1}
&&ds^2 = e^{-\phi} ds^2_{0123} + e^\phi ds_6^2,\nonumber\\
&& {\cal F}_3 = e^{2\phi} \ast_6 d\left(e^{-2\phi} J\right), \nd
where $\phi$ is the dilaton and the Hodge star and the fundamental form $J$ are wrt to the dilaton deformed metric $e^{2\phi} ds_2^6$. The five-brane charge in \eqref{sugra1}
decomposes as \eqref{sourcech} once we express it as a seven-form ${\cal F}_7 = \ast_{10} {\cal F}_3$. The metric $ds_6^2$ is in general a 
noncompact non-K\"ahler
metric that may not even have an integrable complex structure.

If we allow for background three-forms ${\cal F}_3$ and ${\cal H}_3$, the above background \eqref{sugra1} changes. One way to see the change would be to work out the 
precise EOMs. However there exists another way, using a series of duality transformations, to study the background in the presence of the three-form fluxes. The steps 
have been elaborated in \cite{MM, fangpaul, DEM}. The solutions we will study here are specific realizations of the general solutions found and analyzed in \cite{DEM}, where supersymmetry of the final 'dualized' solution was explicitly confirmed\footnote{In addition, the fact that the T-duality transformations lead to solutions that solve explicitly the supergravity EOMs has been shown earlier in \cite{bergkallosh, BHO, bergkal}. In \cite{Gurrieri:2007jg} and \cite{DEM}, this was confirmed using torsion classes. The subtlety that such transformations {\it do not} lead to non-trivial Jacobians follows from the fact that the supergravity fields have no dependence on the T-duality directions. If the supergravity fields start to depend on the T-duality directions, there will arise non-trivial Jacobians as discussed in some details in \cite{gregm}. We thank the referee for raising this question.}.  The idea is to:

\vskip.1in

\noindent $\bullet$ Compactify the spatial coordinates $x^{1, 2, 3}$ and T-dualize three times along these directions. The resulting picture will now be in type IIA theory.

\vskip.1in

\noindent $\bullet$ Lift the type IIA configuration to M-theory and make a boost along the eleventh direction using a boost parameter $\beta$. This boosting will create the
necessary gauge charges.

\vskip.1in

\noindent $\bullet$ Reduce this down to type IIA and T-dualize three times along the spatial coordinates to go to type IIB theory. The IIB background now automatically has the 
three-form fluxes, as well as a five-form flux.

\vskip.1in

\noindent The result of this duality procedure is that the type IIB background \eqref{sugra1} now converts to exactly what we expect in \eqref{back}, 
namely\footnote{There is some subtlety in interpreting the final background with fluxes or with sources. This has been discussed in \cite{fangpaul} which the readers may refer to for details.}:
\bg\label{sugra2}
ds^2  = {1\over \sqrt{h}} ds^2_{0123} + \sqrt{h} ~ds^2_6
= {1\over e^{2\phi/3} \sqrt{e^{2\phi/3} + \Delta}} ~ds^2_{0123} + e^{2\phi/3} \sqrt{e^{2\phi/3} + \Delta} ~ds^2_6, \nd
confirming the low-energy effective D3-brane behavior, and the following background for the three- and the five-form fluxes:
\bg\label{fluxess}
&& {\cal F}_3 = {\rm cosh}~\beta~e^{2\phi} \ast_6 d\left(e^{-2\phi} J\right), ~~{\cal H}_3 = -{\rm sinh}~\beta ~d\left(e^{-2\phi} J\right) ,\nonumber\\
&&d {\cal \widetilde{F}}_5 = - {\rm sinh}~\beta~{\rm cosh}~\beta~ e^{2\phi}~d\left(e^{-2\phi} J\right) \wedge \ast_6 d\left(e^{-2\phi} J\right), \nd
with the type IIB dilaton $e^{\phi_B} = e^{-\phi}$. One may verify that \eqref{sugra2} and \eqref{fluxess} together solve the type IIB EOMs.

We will concentrate on a specific background given by a (generically non-K\"ahler) singular, resolved or deformed conifold. 
The typical internal metric $ds^2_6$ in this class is given by a variant of \eqref{resolve} as:
\bg\label{ds6}
ds^2_6 &=& F_1~ dr^2 + F_2 \left(d\psi + {\rm cos}~\theta_1~d\phi_1 + {\rm cos}~\theta_2~d\phi_2\right)^2 + \sum_{i=1}^2 F_{2+i} \left(d\theta_i^2 + \sin^2\theta_i~d\phi^2\right) \\
&+& F_5 ~\sin~\psi\left(d\phi_1~ d\theta_2 ~\sin~\theta_1 + d\phi_2 ~d\theta_1 ~\sin~\theta_2\right) 
+ F_6 ~\cos~\psi \left(d\theta_1~ d\theta_2 - d\phi_1 ~d\phi_2~\sin~\theta_1~\sin~\theta_2\right), \nonumber \nd
where $F_i(r)$ are warp factors that are functions of the radial coordinate $r$ only\footnote{One may generalize this to make the warp factors $F_i$ functions of 
all coordinates except ($\theta_1, \phi_1$), i.e the directions of the wrapped brane. We will not discuss the generalization here.} and in the following, unless mentioned otherwise, we will only 
consider the resolved conifold, i.e we take $F_5 = F_6 = 0$ henceforth. 
The above background \eqref{ds6} can be easily converted to a background with both ${\cal H}_3$ and ${\cal F}_3$ fluxes by
the series of duality specified above. Using \eqref{sugra2}, our background becomes:
\bg\label{iibform}
&&ds^2 = {1\over e^{2\phi/3} \sqrt{e^{2\phi/3} + \Delta}} ~ds^2_{0123} + e^{2\phi/3} \sqrt{e^{2\phi/3} + \Delta} ~ds^2_6 , \\
&& {\cal F}_3 = - e^{2\phi} {\rm cosh}~\beta \sqrt{F_2\over F_1}\left(g_1 ~e_\psi \wedge e_{\theta_1} \wedge e_{\phi_1}  +
g_2~e_\psi \wedge e_{\theta_2} \wedge e_{\phi_2}\right)\nonumber , \\ 
&& {\widetilde{\cal F}}_5 = -{\rm sinh}~\beta~{\rm cosh}~\beta\left(1 + \ast_{10}\right) {\cal C}_5(r)~d\psi \wedge \prod_{i=1}^2~\sin~\theta_i~ d\theta_i \wedge d\phi_i\nonumber , \\
&& {\cal H}_3 = {\rm sinh}~\beta \Big[\left(\sqrt{F_1 F_2} - F_{3r}\right) e_r \wedge e_{\theta_1} \wedge e_{\phi_1} 
+ \left(\sqrt{F_1 F_2} - F_{4r}\right) e_r \wedge e_{\theta_2} \wedge e_{\phi_2}\Big] \nonumber , \nd
with a dilaton $e^{\phi_B} = e^{-\phi}$ and with $\Delta$ defined as in \eqref{Delta},
\bg\label{linfri}
\Delta = {\rm sinh}^2\beta \left(e^{2\phi/3} - e^{-4\phi/3}\right) , \nd
and $\beta$ is the boost parameter discussed above while
the others, namely ($g_1, g_2, {\cal C}_5$)  are given by:
\bg\label{jmar}
&&g_1(r) = F_3\left(\sqrt{F_1 F_2} - F_{4r} \over F_4\right), ~~~g_2(r) = F_4\left(\sqrt{F_1 F_2} - F_{3r} \over F_3\right),\\ 
&&{\cal C}_5(r) = \int^r~{e^{2\phi} F_3 F_4 \sqrt{F_1 F_2}\over F_1}\left[\left({\sqrt{F_1F_2} - F_{3r}\over F_3}\right)^2 
+ \left({\sqrt{F_1F_2} - F_{4r}\over F_4}\right)^2\right] dr .\nonumber\nd
The above background for the D3-brane is consistent as long as the energy is less than the inverse size of the sphere parametrized by ($\theta_1, \phi_1$). For vanishing
size of the sphere, which would happen for a singular conifold, our analysis continues to hold to arbitrary energies. 

Equation \eqref{iibform} contains all the information that we need, so now the relevant question is to find appropriate warp-factors that allow us to have a non-K\"ahler resolved 
conifold with an integrable complex structure. A simple analysis of the fluxes along the lines of \cite{DEM} will tell us that an integrable complex structure is possible when the dilaton has
no profile in the internal direction. This means 
we can take, without any loss of generality, a vanishing dilaton inducing the 
following complex structure on the internal space:
\bg\label{chela}
\tau_k \equiv (i~{\rm coth}~\beta, i, i). \nd
The metric on the internal space now is not too hard to find if one takes care of all the subtleties pointed out in \cite{DEM}. The subtleties are generically related to flux quantization and 
integrability conditions. Once the dust settles the metric becomes:
\bg\label{aleta}
ds^2 & = & 4 F_{2r}^2 \left({1-G\over 2 + F_2}\right) dr^2 + F_2 (d\psi + {\rm cos}~\theta_1 d\phi_1 + {\rm cos}~\theta_2 d\phi_2)^2 \nonumber\\
& + & G (d\theta_1^2 + \sin^2\theta_1 d\phi_1^2) + G(1-G)\left({F_2\over 2 + F_2}\right) (d\theta_2^2 + \sin^2\theta_2 d\phi_2^2), \nd 
where $F_2(r)$ is taken to be dimensionless. This means all terms of the metric are dimensionless, and thus if $r$ has a dimension of length, the warp-factor should have inverse length dimension. This 
works out fine because the coefficient of $dr^2$ is indeed the derivative of $F_2$. We could also rewrite the metric with dimensionful warp-factors but this would not change any of the physics. 
Note also that $G(r)$ appearing in \eqref{aleta} is not an independent function, but depends on $F_2$ in the following way:
\bg\label{alujen}
(1- G)^3 = {(2+F_2)^3\over F_2(3 + 2F_2)^2}, \nd
and therefore an appropriate choice of $F_2$ will fix the functional form for $G$. Furthermore, the resolution parameter for the resolved conifold is no longer a constant, but a function of the radial
coordinate $r$ that takes the following form:
\bg\label{resa2}
a^2(r) ~ \equiv ~ {(2+ GF_2)G\over 2+ F_2}, \nd
which is by construction a positive definite function provided $G$ remains positive definite everywhere. 
It is definitely a well-behaved function at any point in $r$ since $F_2 > 0$ and if $F_2$ is chosen to be a well-behaved function of $r$. Positivity of $G$ implies that at any point in $r$, 
$F_2$ should satisfy:
\bg\label{lenrae}
F_2^3 + 2 F_2^2 - F_2 ~ > ~ {8\over 3}, \nd
which is not hard to satisfy. This also imples $G < 1$ at any point in $r$. 
A simple choice of $F_2(r)$ would be to consider the following functional form that should make all the warp-factors positive definite:
\bg\label{lenme}
F_2(r) ~ = ~ 1.1022 + \widetilde{F}_2^2(r), \nd
assuming $\widetilde{F}_2(r)$ never hits zero at any point in $r$.
We can also bring our metric \eqref{aleta} to the form \eqref{churamont} by appropriately defining $\delta F, a_1(r)$ and $a_2(r)$. 

It is now time to determine the fluxes that preserve the background supersymmetry. As is well known, the fluxes should be ISD and primitive, so the appropriate choice is to take 
(2, 1) forms. This can be easily worked out from \eqref{iibform}, and 
once we fix the complex structure to be \eqref{chela}, and with the above warp factors and dilaton, the three-form flux takes  particularly simple form \footnote{ where the $E_i$ are defined as:
\begin{equation}\label{c1fom}
E_1 = e_1 + i \coth \beta \, e_2, ~~~~~~~ E_2 = e_3 + i e_4, ~~~~~~~ E_3 = e_5 + i e_6 ,\nonumber\end{equation}  with
\begin{eqnarray}\label{vielbeins}
&& e_1 = \sqrt{F_1\sqrt{H}}e_r, ~~~~~~ e_2 = \sqrt{F_2\sqrt{H}}(d\psi + \cos~\theta_1 d\phi_1 + \cos~\theta_2 d\phi_2) = \sqrt{F_2\sqrt{H}} e_\psi , \nonumber\\
&& e_3 = \sqrt{F_3\sqrt{H}}\left(-\sin~{\psi\over 2} ~e_{\phi_1} + \cos~{\psi\over 2}~e_{\theta_1}\right), 
~~e_4 = \sqrt{F_3\sqrt{H}}\left(\cos~{\psi\over 2} ~e_{\phi_1} + \sin~{\psi\over 2}~e_{\theta_1}\right) , \nonumber\\   
&& e_5 = \sqrt{F_4\sqrt{H}}\left(-\sin~{\psi\over 2} ~e_{\phi_2} + \cos~{\psi\over 2}~e_{\theta_2}\right), 
~~e_6 = \sqrt{F_4\sqrt{H}}\left(\cos~{\psi\over 2} ~e_{\phi_2} + \sin~{\psi\over 2}~e_{\theta_2}\right) , \nonumber\end{eqnarray} }:
\bg\label{G3isdSmallBeta}
{\cal G}_3 & = & {{\rm sinh}~\beta\over 4\sqrt{H} \sqrt{F_1\sqrt{H}}}\left[\left({\sqrt{F_1F_2} - F_{3r}\over F_3}\right) - \left({\sqrt{F_1F_2} - F_{4r}\over F_4}\right)\right]
\left(E_1 \wedge E_3 \wedge {\overline{E}_3} - E_1 \wedge E_2 \wedge {\overline{E}_2}\right) , \\
& = &  {\sqrt{F} \left(2 - F \delta F\right) \over 8 ~{\rm cosech}~ \beta}
\left[\left({r F \delta F -12 a_1 a_{1r}\over r^2 + 6 a_1^2}\right) - \left({r F \delta F -12 a_2 a_{2r}\over r^2 + 6 a_2^2}\right)\right]   
\left(E_1 \wedge E_3 \wedge {\overline{E}_3} - E_1 \wedge E_2 \wedge {\overline{E}_2}\right) , \nonumber
\nd
which is ISD, primitive, and a $(2,1)$ form. In the second line we have used the ansatze \eqref{churamont} with vanishing dilaton. Note also that the three functions $\delta F$, $a_1$, and $a_2$ are constrained by supersymmetry, via \eqref{susycondition}, which is a first order ODE. The SUSY condition also forces the (1, 2) components of $\mathcal{G}_3$ to vanish identically.

One can see that the boost parameter $\beta$, which counts the units of ${\cal F}_3$ flux, or equivalently the number of delocalized (\cite{Gaillard:2008wt}) five-branes, 
in the resolved conifold background, controls the amount of ISD flux. Naively, 
if we take $\beta\rightarrow 0$, the flux vanishes. However the complex structure \eqref{chela} also blows up in this limit, so vanishing $\beta$ case has to be studied differently. This is indeed the
case because, in the language of \cite{DEM}, taking $\beta \to 0$ takes us to the ``before duality'' picture where only RR three-form fluxes are present. Therefore the way we derived our background, we can take $\beta$ arbitrarily small but not 
zero.

This completes our analysis of the supersymmetric fluxes on a non-K\"ahler resolved conifold bacground that allows an integrable complex structure. 
In the following section we will insert a ${\overline{\rm D3}}$-brane in this background and study the fluxes and the corresponding supersymmetry breaking scenario using the world-volume 
action. We start with the bosonic action for a ${\overline{\rm D3}}$-brane in this background.

\subsection{Bosonic action for a $\overline{\rm D3}$-brane}

Before considering a $\overline{\rm D3}$, let us consider a D3. In the previous section we saw how to incorporate the backreaction of a single (or generically $N$) {\it effective} D3-branes in flux background. 
We can compute the bosonic action 
of the D3-brane in this background, not as a probe, but as an actual backreacted object. This is {\it different} from what has been done earlier in 
\cite{grana, marolf1, marolf2, trivedi, martucci, bergkallosh, shiu} where the D3-brane has been considered as a probe in a GKP background \cite{DRS, GKP} of the form:
\bg\label{gkp}
&&ds^2 = e^{2A} g_{\mu\nu} dx^\mu dx^\nu + e^{-2A}g_{mn} dy^m dy^n,\nonumber\\
&& {\cal G}_3 = {\cal F}_3 + \tau ~{\cal H}_3, ~~~{\cal F}_5 = (1 + \ast_{10}) d\alpha \wedge d{\rm vol}_{R^{3, 1}}, \nd
where $\tau = C_0 + i e^{-\phi_B}$ and $\alpha = e^{4A}$. For our case, with the backreaction of the D3-branes taken into account, we can define the following 
quantities:
\bg\label{AO+}
&&e^{2A} = \sqrt{\alpha} = {1\over e^{2\phi/3} \sqrt{e^{2\phi/3} + \Delta}}, ~~g_{\mu\nu} = \eta_{\mu\nu} ,\nonumber\\
&& \Phi_+ = {2\over e^{2\phi}~{\rm cosh}^2\beta - {\rm sinh}^2\beta}, ~~~~~ \Phi_- = 0. \nd
The above equation implies that the scalar fields on a D3-brane are completely massless (as the masses of the scalar fields are determined by $\Phi_-$ \cite{shiu}). Other
details regarding the action can be worked out from \cite{grana, marolf1, marolf2, trivedi, martucci, bergkallosh, shiu}. 

Let us now consider a ${\overline{\rm D3}}$ in this background. We will take this as a probe so that the backreaction of the anti-brane will not be felt strongly in \eqref{iibform}. Details of this
will be discussed in the next section. For the time being we shall assume that a small profile for the dilaton is now switched on, along with small changes in the three-form fluxes. 
Furthermore, the tachyonic instability of the anti-brane will not be visible in the probe limit. The world-volume multiplet on the anti-brane will have the 
usual vector field $A_\mu$ and six scalars $\varphi^m$ associated with the six internal directions of the resolved conifold \eqref{ds6}. The bosonic action in the 
Einstein frame is 
then given by:
\bg\label{bosact}
S_{{\overline{\rm D3}}}
 = - \tau_{D3} l_s^4 
 \int d^4 x\left({\pi \over 2 g_s^2} f_{\mu\nu}f^{\mu\nu} + {\pi\over g_s} g_{mn} {\cal D}_\mu \varphi^m {\cal D}^\mu \varphi^n 
+ {\pi\over g_s} \partial_m \partial_n \Phi_+ \varphi^m \varphi^n  + {\cal L}_{int}\right), \nd
where the interaction lagrangian ${\cal L}_{int}$ is given by the following expression:
\bg\label{lint}
{\cal L}_{int} = {2\pi \over l_s^2 g_s} \partial_m \Phi_+ \varphi^m + {i\pi \over 12} \Phi_+ \left({\rm Re}~G_+\right)_{mnp} \varphi^m \varphi^n \varphi^p 
+ {\pi \over l_s^4 g_s}\Phi_+ , \nd
with $g_{mn}$ to be the metric of the internal non-K\"ahler resolved conifold \eqref{ds6} 
and $G_+ = (\ast_6 + i)G_3$ where $\ast_6$ is the Hodge star with respect to the warped
metric \eqref{gkp}. 

For a conifold background, there are five compact scalars, namely: ($\varphi^{\theta_1}, \varphi^{\phi_1}, \varphi^{\theta_2}, \varphi^{\phi_2}, 
\varphi^{\psi}$), and one non-compact scalar $\varphi^{r}$. The compact scalars are all massless, and the mass of the non-compact scalar is given by:
\bg\label{mass}
m^2_{\varphi^r} &=& {\pi\over g_s}\left({\partial^2 \Phi_+\over \partial r^2}\right)\\ 
&=& {8\pi e^{2\phi}~{\rm cosh}^2 \beta\over g_s\left(e^{2\phi}~{\rm cosh}^2\beta - {\rm sinh}^2\beta\right)^2}
\left[\left({e^{2\phi}~{\rm cosh}^2\beta + {\rm sinh}^2\beta \over e^{2\phi}~{\rm cosh}^2\beta - {\rm sinh}^2\beta}\right)
\left({\partial \phi\over \partial r}\right)^2
-{1\over 2}{\partial^2 \phi \over \partial r^2}\right],\nonumber  \nd
where due to the presence of the linear interaction in \eqref{bosact}, the non-compact scalar is shifted from its original value $\varphi^r$ to the following:
\bg\label{shift}
\widetilde{\varphi}^r \equiv \varphi^r + {1\over l_s^2}\left[{\partial \over \partial r}{\rm log}\left({\partial \Phi_+\over \partial r}\right)\right]^{-1}.\nd
In a generic setting, where the warp-factors and  the dilaton $\phi$ are functions of all the internal coordinates, all the six-scalars would be 
massive and the anti-brane will be fixed at a point in the internal space where the mass matrix is extremised.

However, the background we have constructed has a \emph{constant} dilaton, and thus $\Phi_{+}$ is constant and $\varphi^r$ is massless. If one allows for a small dilaton profile, for example by perturbing beyond the probe limit, a mass is generated for $\varphi^r$. In the limit where $\beta$ is small, this
happens at the point where the dilaton satisfies the following differential equation:
\bg\label{dileq}
{\partial^3\phi \over \partial r^3} - 6\left[{\partial^2\phi \over \partial r^2} - {2\over 3}\left({\partial \phi \over \partial r}\right)^2\right]
{\partial \phi \over \partial r}  + {\cal O}(\beta) = 0.\nd
For the solution discussed above, and allowing for some $\overline{\rm D3}$ backreaction in the form of a small profile for the dilaton, $\Phi_+$ takes the following simple form (for arbitrary values of $\beta$):
\begin{equation}
\Phi_+ = 2 - 4\phi(r)~{\rm cosh}^2\beta + \mathcal{O}(\phi^2).
\end{equation}
This form of $\Phi_+$ will fix $\varphi^r$ to be $0$. The remaining scalars can be stabilized along the lines of \cite{Aharony:2005ez}; the angular moduli recieve masses upon `glueing' the non-compact throat geometry on to a compact Calabi-Yau. Alternatively, one can place the ${\overline{\rm D3}}$ directly on an orientifold plane, as in \cite{Kallosh:2015nia}, which fixes all the scalars and gauge fields\footnote{ For more details on orientifolding conifolds  see   \cite{Park:1999ep, Retolaza:2016alb}, and for the consistency of placing anti-branes on orientifolds of conifolds see \cite{Kallosh:2015nia}. }.

\subsection{SUSY breaking and the fermionic action for a ${\overline{\rm D3}}$ }
\label{sec:backreaction}
Now let us return to the fermionic action, which we gave in equation \eqref{antiD3fermionaction}. The masses of the fermions are dictated by ISD three-form flux ${\cal G}_3$ given in \eqref{G3isdSmallBeta}, which is valid strictly in the probe approximation. The backreaction of the $\overline{\rm D3}$ induces corrections to the flux, which we will come back to shortly.

Staying within the probe approximation, the flux is given by equation (\ref{G3isdSmallBeta}),
\bg\label{daku}
{\cal G}_3 = {\sqrt{F} \left(2 - F \delta F\right)\over 8~ {\rm cosech}~ \beta}
\left[\left({r F \delta F -12 a_1 a_{1r}\over r^2 + 6 a_1^2}\right) - \left({r F \delta F -12 a_2 a_{2r}\over r^2 + 6 a_2^2}\right)\right]   
\left(E_1 \wedge E_3 \wedge {\overline{E}_3} - E_1 \wedge E_2 \wedge {\overline{E}_2}\right) . \nonumber
\nd
Clearly the masses $m_0$ and $m_i$ will be zero (since ${\cal G}_3$ is ISD and primitive). The breaking of supersymmetry is done purely through the mass matrix $m_{ij}$, defined in equation \eqref {eq:mtrip}. Evaluating these masses explicitly, we find
 \begin{equation}\label{massu}
 m_{23} = m_{32} = \frac{\sqrt{2}}{8} i|\mathcal{G}_3| \;\;,\;\; m_{12}=m_{21}=m_{13}=m_{31}=0 ,
 \end{equation}
where $|\mathcal{G}_3|$ is
\begin{equation}
|\mathcal{G}_3| =   {\sinh \beta \over 8}{\sqrt{F} \left(2 - F \delta F\right) }
\left[\left({r F \delta F -12 a_1 a_{1r}\over r^2 + 6 a_1^2}\right) - \left({r F \delta F -12 a_2 a_{2r}\over r^2 + 6 a_2^2}\right)\right]  .
\end{equation}
From this we see that the $\lambda^{2}$ and $\lambda^3$ fermions will have a mass induced by $\mathcal{G}_3$, which spontaneously breaks the $\mathcal{N}=1$ supersymmetry of the resolved conifold. This leaves \emph{two} massless fermions, $\lambda^0$ and $\lambda^1$, as the low energy field content. This is in contrast to an $\overline{\rm D3}$ in a GKP background, studied in \cite{wrase2}, where there was only a single massless fermion. Interestingly, the scale of SUSY breaking is controlled by 
$\delta F(r)$, $a_{1r}$, and $a_{2r}$, and thus we can easily allow for soft breaking of supersymmetry.

\subsection{Perturbing away from the probe limit}

Let us now consider perturbing away from the probe limit, which corresponds to taking the $\overline{\rm D3}$ to be a large yet finite distance away from the D5-brane (fractional D3). We will neglect subtleties regarding boundary conditions, which can lead to divergences in the fluxes when a stack of $\overline{\rm D3}$'s is considered, see e.g.  \cite{Bena:2012bk} and more recently  \cite{Cohen-Maldonado:2015ssa}, and also continue to study only a single $\overline{\rm D3}$. As we will see, even with this issue neglected, backreaction changes the story considerably. In the presence of a probe ${\overline{\rm D3}}$, the background changes from what we have thus far studied. The question then is to compute the changes in the background metric and fluxes to account for the fermionic masses on the anti-brane world-volume. We will \emph{not} attempt to find an exact backreacted solution with an ${\overline{\rm D3}}$, but rather take on a simpler task; we can compute the leading corrections to the fluxes and thus fermion masses by perturbing away from the probe limit.

The situation is not as hard as it sounds. Due to the (perturbatively) probe nature of the ${\overline{\rm D3}}$, and as we hinted before, the tachyonic degree of freedom will not be visible at the 
supergravity level. Furthermore the backreaction of the ${\overline{\rm D3}}$-brane will appear from its energy-momentum tensor that comes solely from the Born-Infeld part (the Chern-Simons 
piece, that can distinguish between a brane and an anti-brane, does not contribute to the energy-momentum tensor). This is good because then at the supergravity level we are effectively inserting 
a three-brane in a wrapped D5-brane background. To compensate for this new source of energy-momentum tensor the warp-factors change slightly as:
\bg\label{fan4}
F_i ~\to ~ F_i + \delta F_i, \nd
where this change is over and above the $\delta F$ change in \eqref{churamont} that was there in the absence of ${\overline{\rm D3}}$-brane\footnote{Note that 
due to the probe nature, $\delta F_5 = \delta F_6 = 0$ along with vanishing ($F_5, F_6$), so that the {\it form} of the metric remains \eqref{ds6} and the topology doesn't change.}. 
The dilaton $\phi$ also changes from zero to $\delta\phi$,
but, as a first trial, we keep the complex structure of the non-K\"ahler resolved conifold fixed to \eqref{chela} (as we shall see, this will have to be changed).  
Note that for a supersymmetric perturbation, the complex structure would have also changed exactly in a way so as 
to remove any ($1, 2$) fluxes.  Taking this into account, the ISD primitive ($2, 1$) flux \eqref{G3isdSmallBeta} now changes to the following additional piece:
\bg\label{21change}
\delta{\cal G}_3^{(1)} & = & {{\rm sinh}~\beta\over 4\sqrt{H} \sqrt{F_1\sqrt{H}}} \left(1 + {\delta F_1\over 2 F_1} + {3\over 4}{\delta H\over H}\right) 
\Bigg[{\sqrt{F_1F_2}\over 2}\left({\delta F_1\over F_1} + {\delta F_2\over F_2}\right) \left({1\over F_3} - {1\over F_4}\right) + \left({\delta F_{4r}\over F_4} - {\delta F_{3r}\over F_3}\right)\\
 & + & \left({\sqrt{F_1F_2} - F_{4r}\over F_4}\right) \left({\delta F_4\over F_4} -\delta\phi\right)  - 
\left({\sqrt{F_1F_2} - F_{3r}\over F_3}\right) \left({\delta F_3\over F_3} -\delta\phi\right)\Bigg] \left(E_1 \wedge E_3 \wedge {\overline{E}_3}  - E_1 \wedge E_2 \wedge {\overline{E}_2}\right),
\nonumber \nd  
which is again a primitive ($2, 1$) form. When combined with the primitive ($2, 1$) piece that we had in \eqref{G3isdSmallBeta}, this would enter the mass formula given in \eqref{eq:mtrip} to give masses to the corresponding fermions. 
Note that, the $E_i$'s appearing above are the {\it original} vielbein used earlier to write the (2, 1) flux \eqref{G3isdSmallBeta}, but could be replaced by the 
modified vielbein
under \eqref{fan4}, i.e:
\bg\label{modviel}
E_i ~ \to ~ E_i + \delta E_i, \nd 
without changing any physics. 
This will be also be the case for all other (2, 1) and  
(1, 2) perturbations that we shall discuss below: we will express them in terms of old vielbeins although we could also use \eqref{modviel}. 
Using the old vielbeins $E_i$, we do however develop an
{\it additional} contribution to the (2, 1) flux, other than \eqref{G3isdSmallBeta} and \eqref{21change}, 
that typically takes the following form:
\bg\label{confluxform}
\delta {\cal G}_3^{(2)} = \left(\alpha_1 \delta F + \alpha_2 a_{1r} + \alpha_3 a_{3r}\right) \left(E_i \wedge \delta E_j \wedge \overline{E}_k \pm 
\sigma~ E_i \wedge E_j \wedge \delta\overline{E}_k\right), \nd
where $\alpha_i(r)$ and $\sigma(r)$ are certain well defined functions of $r$ that could be derived from our flux formulae discussed above. We cannot simply ignore 
this term as it is of the same order as the second line in \eqref{21change} above, but we can absorb this in \eqref{G3isdSmallBeta} by resorting to the 
modified veilbein \eqref{modviel}. The conclusion then remains unchanged: all $\delta{\cal G}_3^{(k)}$ will be expressed in terms of $E_i$, but the original (2, 1) 
flux \eqref{G3isdSmallBeta} will now be expressed in terms of \eqref{modviel} under perturbative backreaction of $\overline{\rm D3}$-brane. 

Coming back to our analysis, 
the primitive ($2, 1$) pieces are responsible in determining the masses, but we do also get another (2, 1) piece that is {\it neither} primitive nor ISD. 
This appears because we haven't changed our 
complex structure, and it is given by the following form: 
\bg\label{21nonp}
\delta{\cal G}_3^{(3)}  =  {\cal G}_0^{(\delta\phi)}
\left[\left({\sqrt{F_1F_2}-F_{4r}\over F_4}\right)E_1 \wedge E_2 \wedge {\overline{E}_2} + 
\left({\sqrt{F_1F_2}-F_{3r}\over F_3}\right)E_1 \wedge E_3 \wedge {\overline{E}_3}\right], \nd 
which becomes an ISD primitive form when the sum of the coefficents of the two terms vanish. This is no surprise because it is exactly the supersymmetry condition that we had in \cite{DEM}. We have also
defined the coefficient ${\cal G}(r)$ in terms of the warp-factors $H$ and $F_1$ in the following way:
\bg\label{calgr}
{\cal G}_0^{(\delta\phi)} \equiv -{\delta\phi~{\rm sinh}~\beta\over 4\sqrt{H} \sqrt{F_1\sqrt{H}}} \left(1 + {\delta F_1\over 2 F_1} + {3\over 4}{\delta H\over H}\right).\nd 
Additionally, under supersymmetry $\delta \phi$ vanishes, so this term never shows up. For the present case, clearly we cannot impose the supersymmetry conditions. However if we change the 
complex structure \eqref{chela} a bit in the following way:
\bg\label{ccchange}
\delta\tau_k ~ = ~ (i\delta\phi~{\rm coth}~\beta, 0, 0), \nd
instead of keeping it completely rigid as we discussed above, we can make this term vanish. Note that some care is required to interpret this result. As mentioned earlier, we can change the 
complex structure to absorb any appearance of (1, 2) forms so that supersymmetry is restored. This case should then be interpreted differently. As we shall see below, we do get (1, 2) forms and
they will be non-zero for the shifted complex structure \eqref{ccchange} as well as for the original complex structure \eqref{chela}. 

The (1, 2) piece is given by the following form:
\bg\label{12lalop}
\delta {\cal G}_3^{(4)} & = & {{\rm sinh}~\beta\over 4\sqrt{H} \sqrt{F_1\sqrt{H}}} \left(1 + {\delta F_1\over 2 F_1} + {3\over 4}{\delta H\over H}\right) 
\Bigg[{\sqrt{F_1F_2}\over 2}\left({\delta F_1\over F_1} + {\delta F_2\over F_2}\right) \left({1\over F_3} + {1\over F_4}\right) 
- \left({\delta F_{4r}\over F_4} + {\delta F_{3r}\over F_3}\right)\nonumber\\
 & - & \left({\sqrt{F_1F_2} - F_{4r}\over F_4}\right){\delta F_4\over F_4} - 
\left({\sqrt{F_1F_2} - F_{3r}\over F_3}\right){\delta F_3\over F_3}\Bigg] \left(E_2 \wedge {\overline{E}_1} \wedge {\overline{E}_2}  + E_3 \wedge {\overline{E}_1} \wedge {\overline{E}_3}\right), \nd
which is an ISD but non-primitive form, and therefore breaks supersymmetry. 
As before, we have ignored terms of the form $\delta F_i \delta F_j$ and $\delta F_i \delta \phi$, as we are assuming the perturbations to be small. 
When the perturbations are not small we need to use more exact expressions which can be derived with some effort, but we will not do this here.  The above (1, 2) form \eqref{12lalop} enters the mass formula \eqref{eq:mmix}, inducing a non-zero $m_{\bar{1}}$. This acts as an interaction between $\lambda^{\bar{1}}$ and $\lambda^{\bar{0}}$. Similarly, $\overline{\delta {\cal G}}_3^{(3)} $ induces an interaction $m_1 \lambda^0 \lambda^1$. This is given by
\begin{equation}
m_1 = \frac{1}{\sqrt{2}} e^{\delta\phi} |\overline{\delta {\cal G}}_3^{(3)} | ,
\end{equation}
where $|\overline{\delta {\cal G}}_3^{(3)} |$ is the coefficient of $\left(E_2 \wedge {\overline{E}_1} \wedge {\overline{E}_2}  + E_3 \wedge {\overline{E}_1} \wedge {\overline{E}_3}\right)$ in equation \eqref{12lalop}.  

Note that in deriving the perturbations to our background we did not find any (0, 3) or IASD forms. 
This is expected from the probe nature of our analysis. On the other hand the (1, 2) form that we got above
in \eqref{12lalop} cannot be absorbed by the change in the complex structure \eqref{ccchange}. However one might ask if a more generic analysis could be performed. In other words, is it possible to 
find the most generic (2, 1) and (1, 2) perturbations in the non-K\"ahler resolved conifold background? 

The way to answer this question would be to first find the complete basis for the (2, 1) and (1, 2) forms in the resolved conifold background. This has been studied in \cite{sully}, and we 
reproduce it here for completeness. The basis for the (2, 1) forms are:
\bg\label{21basis}
&& u_1 \equiv E_1 \wedge E_2 \wedge {\overline{E}_2} - E_1 \wedge E_3 \wedge {\overline{E}_3}, ~~~~~ 
u_2 \equiv E_1 \wedge E_2 \wedge {\overline{E}_3} - E_1 \wedge E_3 \wedge {\overline{E}_2} \nonumber\\ 
&& u_3 \equiv E_1 \wedge E_2 \wedge {\overline{E}_1} + E_2 \wedge E_3 \wedge {\overline{E}_3}, ~~~~~
u_4 \equiv E_1 \wedge E_3 \wedge {\overline{E}_1} - E_2 \wedge E_3 \wedge {\overline{E}_2} \nonumber\\ 
&& u_5 \equiv E_2 \wedge E_3 \wedge {\overline{E}_1}, \nd
where all of them are ISD and primitive. The first basis, 
$u_1$, was used earlier to write both the original and the perturbed (2, 1) forms. The bases ($u_2, ..., u_5$) are useful when the ${\overline{\rm D3}}$ backreaction is not as simple as
\eqref{fan4}. Thus a generic (2, 1) perturbation can be of the form:
\bg\label{genpert}
\delta {\cal G}_3^{(2, 1)} ~ = ~ \sum_{n = 1}^5 a_n u_n, \nd
where $a_n$ could be functions of all the coordinates of the internal non-K\"ahler resolved conifold. We can then use \eqref{genpert} in \eqref{eq:mtrip} to expresses the masses of the relevant 
fermions on the ${\overline{\rm D3}}$-brane.  Most importantly, it will in general no longer be the case that $\lambda^1$ is massless, since more general $(2,1)$ fluxes induces non-zero masses, i.e. we will now have
\begin{equation}
m_{12}\neq 0, \;\;\;\;m_{13}\neq0.
\end{equation}   
One may similarly construct the complete basis for the (1, 2) forms for the resolved conifold background. We will again require our basis forms to be ISD to solve the background EOMs. For a (1, 2) 
form this is possible only if it is proportional to the fundamental form $J$, thus restricting the number of such forms to be just three. They are given by \cite{sully}:
\bg\label{12basis}
&& w_1 \equiv E_1 \wedge {\overline{E}_1} \wedge {\overline{E}_3} + E_2 \wedge {\overline{E}_2} \wedge {\overline{E}_3}, ~~~~~
w_2 \equiv E_1 \wedge {\overline{E}_1} \wedge {\overline{E}_2} - E_3 \wedge {\overline{E}_2} \wedge {\overline{E}_3} ,\nonumber\\  
&& w_3 \equiv E_2 \wedge {\overline{E}_1} \wedge {\overline{E}_2} + E_3 \wedge {\overline{E}_1} \wedge {\overline{E}_3}, \nd  
where one may check that they are ISD but not primitive. We had used $w_3$ earlier to express the (1, 2) perturbation in \eqref{12lalop}. Thus a more generic non-supersymmetric perturbation 
in the presence of a ${\overline{\rm D3}}$-brane can be expressed by the following (1, 2) form:
\bg\label{lentuba}
\delta{\cal G}_3^{(1, 2)} = \sum_{n = 1}^3 b_n w_n, \nd
where $b_n$, as for $a_n$ above, could be generic functions of all the coordinates of the internal non-K\"ahler resolved conifold. This could now be inserted into \eqref{eq:mmix} to determine
the mixing between the $\lambda_{\pm}^0$ and $\lambda_{\pm}^i$ fermions, i.e.
\begin{equation}
m_{1}\neq0.
\end{equation}  
The consequence of this is that the backreaction-induced fluxes give a mass to $\lambda^0$ and $\lambda^1$, and hence there are \emph{no massless fermions left in the spectrum}.  This is a striking difference to the probe approximation, where there were two massless fermions. 

{Let us take a moment to consider why this is the case. From the supergravity perspective, a $\overline{\rm D3}$ is equivalent to a D3. The background we are considering has a wrapped D5-brane, and since a D3-D5 system is non-supersymmetric, the induced fluxes will include supersymmetry breaking fluxes. It is these fluxes which give a mass to the would-be massless fermions on the $\overline{\rm D3}$ worldvolume. In the GKP analysis of \cite{wrase2}, there was no D5-brane, and thus this issue will not arise when considering backreaction.}

This completes our discussion of spontaneous supersymmetry breaking via massive fermions on the ${\overline{\rm D3}}$-brane world-volume. In the following section we will briefly dwell on certain aspects of moduli stabilization and de Sitter space.  

\subsection{Moduli stabilization and de Sitter vacua}

In order to construct a concrete phenomelogical model, the resolved
conifold geometry we have studied should be {\it glued} on to a compact, non-K\"ahler space. As discussed in \cite{Aharony:2005ez}, and also \cite{Kenton:2014gma}, this glueing induces corrections to the ${\overline{\rm D3}}$ scalar moduli masses.

In addition to this, a compact space requires charge cancellation. Since charge cancellation is a global requirement, the necessary fluxes can be placed far from the resolved conifold which contains the ${\overline{\rm D3}}$, so as not to disrupt the local dynamics we have studied. In other words, for the case that we study here, the internal six-dimensional manifold 
\eqref{ds6} should
be thought of as extending to a fixed radius $r = r_0$, and beyond which a compact manifold is attached. The boundary condition implies that at $r = r_0$, the 
compact manifold should have a topology of ${\bf S}^2 \times {\bf S}^3$.
The compact manifold is equipped with the right amount 
of fluxes etc that is necessary for global charge cancellation.  

Finally, we note that moduli stabilization should be included in to this picture. We need to consider two sets of moduli: the K\"ahler and the complex structure moduli of our non-K\"ahler space. 
The moduli of compactifications on non-K\"ahler manifolds was discussed in \cite{Becker:2003yv}, and reviewed in \cite{Becker:2003dz}. An interesting feature of these models is that 
the radial modulus and the complex structure moduli can be stabilized at tree-level whereas the other K\"ahler moduli, including the axio-dilaton need 
additional non-perturbative effects for stabilization. There are also other moduli, namely the moduli of the ${\overline{\rm D3}}$-brane, fractional three-branes and possible seven-branes (that we 
didn't discuss here, but are nonetheless important).

From the point of view of Einstein equations, the existence of de Sitter vacua is rather non-trivial to see. Switching on \eqref{genpert} and \eqref{lentuba} gives masses to worldvolume fermions and 
simultaneously fixes the complex structure moduli (including the radial modulus) of our non-K\"ahler space. However the potential generated by the susy breaking flux \eqref{lentuba}:
\bg\label{dhakka}
V = {1\over 2 \kappa_{10}^2}\int {\delta{\cal G}_3^{(1, 2)} \wedge \ast {\overline{\delta{\cal G}}}_3^{(1, 2)}\over {\rm Im}~\tau}, \nd
where $\tau$ is the axio-dilaton, vanishes identically.
This means the presence of a ${\overline{\rm D3}}$-brane 
takes a supersymmetric AdS space to a non-supersymmetric one, and therefore doesn't contribute any positive vacuum energy to the system. 
This conclusion is not new and 
is another manifestation of the no-go condition of Gibbons-Maldacena-Nunez \cite{GMN}, recently updated in \cite{Mia}. This means to allow for a  
positive cosmological solution in the four space-time direction, the no-go condition should be 
averted\footnote{All the energy-momentum tensors are computed using both the bosonic and the fermionic terms on the branes and the planes. 
Note that the no-go conditions in \cite{GMN, Mia} were derived exclusively using the bosonic terms on the branes and the planes. However if we use 
\eqref{oneroof} (see section 4) to define the pullbacks of the type IIB fields on the branes and the planes, we can easily see that the conclusions of \cite{GMN, Mia} 
remain unchanged in the presence of the fermionic terms.}. 

This then brings us to the recent study done in \cite{Mia} from an uplift in M-theory. Quantum corrections play an important role, and positive cosmological constant is only achieved in 
four space-time directions if the following condition is satisfied:
\bg\label{chamatkar}
\langle {\cal T}^\mu_\mu \rangle_q ~ >  ~  \langle {\cal T}^m_m \rangle_q, \nd
which is a generalization of the classical condition studied in \cite{GMN}. Here ${\cal T}_{mn}$ is the energy-momentum tensor and the subscript $q$ 
denote the quantum part of it. For more details, and
the derivation of this, the readers may want to refer to \cite{Mia}. 

This indicates that a concrete realization of de Sitter vacua in this context, and a precise connection to KKLT \cite{kklt}, would thus require including at least a subset of the above corrections (similar to `K\"{a}hler Uplifting' \cite{Westphal:2006tn}). Note that our setup would not involve to the KPV process \cite{Kachru:2002gs}, whereby a stack of $\overline{\rm D3}$'s polarize into an NS5, as we are only considering a single anti-brane.

\section{Probe $\overline{\rm D7}$ in a GKP Background}

In the previous section we generalized the work of \cite{wrase1, wrase2} to a more general background, and found several interesting features. We now consider a different generalization: we turn our attention to an $\overline{\rm D7}$ brane in a GKP background. Similar to the $\overline{\rm D3}$ case, the $\overline{\rm D7}$ brane differs from the D7 brane only in the sign of $\kappa$-symmetry projector, and the charge under the RR fields.  The embedding of D7 branes into flux compactifications has been the focus of many works; for example \cite{Ouyang:2003df}, \cite{Chen:2008jj}, \cite{sully}, and \cite{Dymarsky:2009cm}.  In particular, many details of the D7 and $\overline{\rm D7}$ fermionic action were worked out in \cite{luestmar} and \cite{Bandos:2006wb}. 

Placing a $\overline{\rm D7}$ in a warped $\mathcal{N}=1$ background will spontaneously break supersymmetry. {The breaking of supersymmetry manifests itself in the fermionic action via a mass for the fermions} (see \cite{luestmar} for details), {and the spontaneous nature of SUSY breaking can be deduced via the condition discussed in Section \ref{sec:spont}}. Furthermore, for general background fluxes, all the $\overline{\rm D7}$ worldvolume fermions are massive. Only under special circumstances will there remain a massless fermion in the low energy spectrum; demonstrating this will be the focus of this section. We will find that, under suitable conditions, we have not only one massless fermion, but many. This is similar to the the $\overline{\rm D3}$ in a resolved conifold case studied in Section 2, where (in the probe approximation) we found not one but two massless fermions.

\subsection{The fermionic action for a $\overline{\rm D7}$ in a flux background }
\label{app}

The quadratic fermionic action for a single Dp-brane (in string frame) is detailed in \cite{martucci}, we will follow their conventions in what follows. The only difference for an anti-brane is in the $\kappa$-symmetry projector, which flips sign relative to the brane case. For the case of $p=7$ this reads:
\begin{equation}
\label{fermaction}
S^{\overline{\rm D7}}_f = - \frac{1}{2}T_7 (2 \pi \alpha')^2 \int \mathrm{d}^8 \xi \; e^{  \phi} \; \sqrt{- \mbox{det}(G +  \mathcal{F})} \; \bar{\theta} 
\left[1- \Gamma_{\overline{\rm D7}} (\mathcal{F})\right] \left(\slashed{\mathcal{D} }- \Delta \right) \; \theta  ,
\end{equation}
where we scaled our action by an overall factor of $(2 \pi \alpha)'^2$ (to match with the convention of writing the gauge field as 
$2 \pi \alpha' {F}_{\mu \nu}$). As before, the spinor $\theta$ is a 10-dimensional 
64(32) real(complex) component Majorana spinor, which is a doublet of 10-dimensional (left-handed) 32(16) real(complex) component Majorana-Weyl spinors. 

The factor $\left[1- \Gamma_{\overline{\rm D7}} (\mathcal{F})\right]$ is the $\kappa$-symmetry projector, which depends on the worldvolume flux $\mathcal{F}$, and we have defined:
\begin{equation}
\Gamma_{\overline{\rm D7}} = -i \sigma_2 \frac{1}{\sqrt{-g}} \Gamma_{01234567} + \mathcal{O(\mathcal{F})} ,
\end{equation}
and we take the brane to be along the $x^0, ..., x^7$ coordinate directions. The covariant derivative $\widetilde{\mathcal{D}}$ on the brane is defined as:
\begin{equation}
 \slashed{\mathcal{D}} = ({M}^{-1})^{\alpha \beta} \Gamma_\beta \widetilde{D}_{\alpha},
\end{equation}
where ${M}_{ab}$ is defined using ${\cal F}_{ab}$ and the pull-back of the metric $g_{ab}$ as:
\begin{equation}
\label{Mab}
{M}_{a b} =g_{ab} + \mathcal{F}_{ab} ,
\end{equation}
with $\mathcal{F} = P[{\cal B}_{(2)}] + 2 \pi \alpha' F_2$.  We have also defined $\widetilde{D}_{\alpha}$ as a shifted covariant derivative, 
\begin{equation}
\widetilde{D}_m = D_m \mathbb{I}_2 +\sigma_1 W_m  ,
\end{equation}
which we shall define in more detail momentarily. It is important to note that the contraction $\slashed{D} = \Gamma^m D_m$ sums only over the indices on the brane-worldvolume, and as mentioned above, we will take the brane to be oriented along the ($x^0, x^1,... x^7$) directions. In contrast to this, the contractions appearing in $\Delta$ will sum over \emph{all} 
indices\footnote{We take our three-form fluxes to be only in the internal space.}, for example $\Delta$ contains the term $\Gamma^{MNP} {\cal H}_{MNP}$ where $M,N,P= 0..9 $. We can further decompose ${\cal H}_{MNP}$ into pieces with 0, 1, and 2, indices along the transverse two-dimensional space parametrized by ($x^8,x^9$) coordinates.

In a general GKP background the worldvolume flux $\mathcal{F}$ will be non-zero, and this cannot be gauged away. To make our analysis simple, we will focus on a class of backgrounds with the property that ${\cal B}_2$ is constant along the brane worldvolume, i.e. ${\cal B}_2={\cal B}_2(x^8,x^9)$, and there is an equal and opposite DBI gauge $F_{2}$, such that $\mathcal{F}=0$.  This allows us to take the $M_{ab}$ appearing in equation (\ref{Mab}) as simply $g_{ab}$, and $\Gamma_{\overline{\rm D7}}$ to be $ -i \sigma_2 \frac{1}{\sqrt{-g}} \Gamma_{01234567}$. Recall that a GKP background also comes equipped with a self-dual five-form flux $\widetilde{\cal F}_5$, given by
\begin{equation}
\widetilde{\cal F}_5 = (1 + \ast) \left(\mathrm{d}\alpha \wedge \mathrm{d}x^0 \wedge  \mathrm{d}x^1  \wedge \mathrm{d}x^2  \wedge \mathrm{d}x^3 \right),
\end{equation}
where the function $\alpha$ depends on the coordinates of the internal space, and is responsible for setting the profile of the warp factor, i.e. $\alpha = e^{4A}$. We will see that $\widetilde{\cal F}_5$ generically contributes to the fermion masses, unless $\alpha=\alpha(x^8,x^9)$, i.e. $\alpha$ is independent of the brane coordinates.

Let us consider an explicit choice of background flux which realizes this. We again define in the standard way ${\cal G}_3= {\cal F}_3 - \tau {\cal H}_3$. A choice of ${\cal G}_3$ which meets the above criteria is:
\begin{equation}
\label{G3example}
{\cal G}_3 = N \; E_1 \wedge E_2 \wedge \overline{E}_3 ,
\end{equation} 
where $N$ is a constant and we take complex structure $J=(i,i,i)$, i.e. $z^1=x^4 + i x^5$ and so on. One can easily check that this is ISD and primitive\footnote{To avoid clutter we are using the same symbol $E_i$ to denote the vielbeins as before although now the definitions of the vielbeins are very different. Furthermore since the background is no longer a non-K\"ahler resolved conifold we are not restricted to the basis \eqref{12basis} to express the three-form ${\cal G}_3$.}. The corresponding ${\cal B}_2$ and ${\cal C}_2$ which generate this ${\cal G}_3$ are:
\begin{eqnarray}
\label{B2C2}
&& {\cal C}_2 = N \left(x^4 \, {d}x^6 \wedge {d}x^8 - x^4\, {d}x^7 \wedge {d}x^9 - x^5 \,{d}x^6 \wedge {d}x^9 - x^5 \, {d}x^7 \wedge {d}x^8\right) \nonumber\\
&&{\cal B}_2 = N e^{\phi_0} \left(x^9 \, {d}x^4 \wedge {d}x^6 + x^8\, {d}x^4 \wedge {d}x^7 + x^8 \,{d}x^5 \wedge {d}x^6 + x^9 \, {d}x^5 \wedge {d}x^7\right), 
\end{eqnarray}
where we take the dilaton to be constant $\phi=\phi_0$. With the above example in mind, we will proceed in our analysis with a general 
${\cal G}_3$, but with the assumption that $F_2=-P[B_2]$ and hence $\mathcal{F}=0$.

As mentioned above, the IIB spinor $\theta$ is actually a doublet of 16-component left-handed (i.e. same chirality) Majorana-Weyl spinors; this `doublet' is a 32 component Majorana spinor, note that it is \emph{not} Weyl. The gamma matrices in this 64 component representations are related to the 16 component representations by:
\begin{equation}
\Gamma^{\rm doublet}_m = \Gamma_m \otimes \mathbb{I}_2 ,
\end{equation}
as in, e.g. below equation 85 in \cite{martucci}.

We gauge fix $\kappa$-symmetry by enforcing the $\kappa$-symmetry projection to satisfy the following condition, namely:
\begin{equation}
\bar{\theta} \left( 1 + \Gamma_{\overline{\rm D7}} \right) =0.
\end{equation}
This enforces a relation between $\theta_{1,2}$ componenst of the doublet $\theta$, given by:
\begin{equation}
\label{kappacondition}
\theta_2 =  \Gamma_{012...7} \theta_1 .
\end{equation}
This choice of gauge fixing was used in recent papers by Kallosh et al., for example \cite{wrase1, wrase2}, as it is consistent with an orientifold projection. Alternatively, one could use a condition $\theta_2=0$, as was used in papers by Martucci et al., e.g. \cite{martucci} and \cite{marolf1, marolf2}. Here, we will only use the condition 
above, namely,  $\theta_2 =  \Gamma_{012...7} \theta_1$.

Lastly, we note that the operators $W_m$ and $\Delta$ appearing in equation (\ref{fermaction}) are given by (see for example \cite{wrase2}):
\bg
\label{woperator} 
&& \Delta = - \frac{1}{2} \Gamma^M \partial_M \phi  - \frac{1}{24} \left({\cal H}_{MNP} \sigma_3  - e^\phi {\cal F}_{MNP} \sigma_1 \right) \Gamma^{MNP}\\
&& W_m = - \frac{1}{4} e^\phi (i \sigma_2) {\cal F}_m + \frac{1}{8} \left( {\cal H}_{mNP} \sigma_3 - e^\phi {\cal F}_{m NP} \sigma_1 \right) \Gamma^{NP}
 - \frac{1}{8 \cdot 4!} (i \sigma_2)e^\phi {\cal F}_{NPQRS} \Gamma^{NPQRS} \Gamma_m \nonumber
\nd
where $m =4,5,6,7$, and $M,N=0,1,...,9$. Additionally   
any quantity not appearing with a $\sigma_i$ is implicitly a tensor product with the $2\times2$ identity matrix. 

We can now expand our action (\ref{fermaction}), using the operators (\ref{woperator}) and the $\kappa$-symmetry fixing condition (\ref{kappacondition}). We use the fact that the fluxes are only in the internal space, and that the only non-vanishing bilinears for $10d$ Majorana-Weyl spinors have 3 or 7 gamma matrices. The action can be written in terms of $\theta_1$ as:
\begin{equation}
\label{fermactionSimplified}
S^{\overline{\rm D7}}_f = - \frac{1}{2}T_7 (2 \pi \alpha')^2 \int \mathrm{d}^8 \xi \; e^{  \phi} \; \sqrt{- \mbox{det} G} \; \mathcal{L}_{\theta} ,
\end{equation}
where $G$ is the warped metric, and $\mathcal{L}_{\theta}$ is given purely in terms of $\theta_1$ as:
\begin{equation} 
\mathcal{L}_{\theta} = 2 \bar{\theta}_1 \left[ \Gamma^m D_m  - \frac{3}{16}  e^{\phi }\Gamma^{mna} \left({\cal F}_{mna} \Gamma_{012...7} + e^{-\phi} {\cal H}_{mna}\right)-\frac{5}{16} e^{\phi} \Gamma^{mnp} {\cal F}_{0123}^{\;\;\;\;\;\;\;q}\epsilon_{mnpq} \right] \theta_1, 
\end{equation}
with the indices running as $m,n,p,q = 4,5,6,7$ and $a = 8,9$.
Note the interesting feature that the only 3-form fluxes which contribute to the action are those with two-legs along the brane, and one leg transverse to the brane. The other contributions, (1) 3 legs along the brane, 0 transverse and (2) 1 leg along the brane, 2 transverse, cancel out of the action. As we see, there is a possible contribution from the 5-form flux when all legs of the flux lie along the brane. This can be made to vanish if we impose that $\alpha$ depend only on the transverse directions to brane. This is different from the $\mathrm{\overline{D3}}$ case, where the $\widetilde{\cal F}_5$ term simply did not contribute, regardless of the choice of $\alpha$. We will return to this point in Section \ref{sec:F5}; for the moment we will take $\alpha=\alpha(x^8,x^9)$ and hence $\widetilde{\cal F}_5$  will not contribute to the masses. There can generally also be a contribution from the 1-form flux, but a GKP background doesn't have these, due to the lack of 1-cycles on a CY 
manifold\footnote{Note that we are putting a $\overline{\rm D7}$ in a GKP background with a constant dilaton and zero axion. The backreacted 
axionic source of the $\overline{\rm D7}$ is suppressed by $g_s$ and to this order we are not taking this to backreact on the $\overline{\rm D7}$ world-volume (the axion will only be 
along ($x^8, x^9$) directions). This differs slightly in 
spirit of the previous section where due to the non-supersymmetric nature of the D3-D5 system, it was essential to take the perturbative backreactions into account, otherwise 
certain aspects of the physics would not have been visible.}.

The action \eqref{fermactionSimplified} can be simplified further by using $\Gamma_{0...9} \theta_1 = \theta_1$, which implies that ${\cal F}_{mna} \Gamma_{012...7} \theta_1 
= (\ast_2 {\cal F}_3)_{mna} \theta_1$, where $\ast_2$ is hodge duality in the ($x^8, x^9$) directions. We can also write this in terms of the familiar 
$\mathcal{G}_3 = {\cal F}_3 - i e^{-\phi} {\cal H}_3$ along with the following nomenclatures: ISD2 as the ``imaginary self-dual'' {along the transverse two-cycle} and 
IASD2 as the ``imaginary anti-self-dual'' again {along the transverse two cycle} pieces of $\mathcal{G}_3$ as
\begin{equation}
\mathcal{G}_3 = \mathcal{G}_3 ^{\rm IASD2} + \mathcal{G}_3 ^{\rm ISD2}, \;\;\;\; \mathcal{G}_3 ^{\rm ISD2} = \frac{1}{2} \left( \mathcal{G}_3  - i \ast_2 \mathcal{G}_3\right), \;\;\;\; \mathcal{G}_3 ^{\rm IASD2} = \frac{1}{2} \left( \mathcal{G}_3  + i \ast_2 \mathcal{G}_3\right),
\end{equation}
which is equivalent to the decomposition
\begin{equation}
{\cal H}_3 = \frac{i}{2} e^\phi \left(\mathcal{G}_3 - \bar{\mathcal{G}}_3 \right) \;\;,\;\; {\cal F}_3 = \frac{1}{2} \left(\mathcal{G}_3 + \bar{\mathcal{G}}_3 \right).
\end{equation}
With these definitions the action becomes
\begin{equation}
\label{finalD7action}
\mathcal{L}_{\theta} = 2 \bar{\theta}_1 \left[ \Gamma^m D_m - \frac{3 i}{32}  e^{\phi }\Gamma^{mna} \left(  \mathcal{G}_3 ^{\rm ISD2}  - \bar{\mathcal{G}}_3 ^{\rm ISD2} \right)_{mna} \right] \theta_1;  \;\;\;\; (m, n) = 4,5,6,7; \;\;\;\; a = 8,9.
\end{equation}
Thus the worldvolume fermions on the $\overline{\rm D7}$ brane will have masses determined by ISD2 $\mathcal{G}_3$ flux, where the `dual' in ISD2 refers to space \emph{transverse to the brane} (and not the full internal space). For our example $\mathcal{G}_3$ given in equation (\ref{G3example}), the flux is purely ISD2 and thus will contribute to the masses. These masses spontaneously break the background $\mathcal{N}=1$ supersymmetry. 

We could also include flux which is ISD $-$ and thus solves the equations of motion for a GKP background $-$ but which is \emph{not} ISD2, and hence will not contribute to the fermion masses. An example of such a flux is
\begin{equation}
\mathcal{G}_3 = M \left( E_1 \wedge \overline{E}_1  - E_2 \wedge \overline{E}_2  \right) \wedge E_3
\end{equation}
 which is purely IASD2, and thus will not enter equation (\ref{finalD7action}). Such a flux would come from a ${\cal B}_2$ of the form 
\bg\label{sur}
{\cal B}_2 = - M e^{\phi_0} x^9 \cdot \left( \mathrm{d}x^4 \wedge \mathrm{d}x^5  - \mathrm{d}x^6 \wedge \mathrm{d}x^7 \right), \nd 
and a similar form for ${\cal C}_2$.

 \subsection{Fermions in $4d$ and spontaneous SUSY breaking in a GKP background}
\label{D74d}

We can already see that supersymmetry will be spontaneously broken by the $\overline{\rm D7}$ in the presence of three-form fluxes. What remains to be checked is if there remains a massless fermion in the four dimensional effective theory. 

In the absence of $\mathcal{G}_3$ flux, the massless fermions in the $4d$ theory are those who's dependence on the coordinates of the internal 4-cycle wrapped by the brane is harmonic. The exact spectrum of effective $4d$ fermions is therefore given by the cohomology classes of the wrapped cycle. On the other hand the coupling of the $\mathcal{G}_3$ flux to the fermions is governed by the structure of the spinors, so we do not need to know the full details of the topology of the wrapped cycle to know whether some of these fermions remain massless. Indeed, most of our calculation proceeds in the same fashion and certainly in the same spirit as the $\overline{\rm D3}$ case\footnote{Without the (1, 2) perturbations of course.}. 

The 16 component spinor $\theta_1$ can decomposed into two 8 component spinors $\theta_{1+}$ and $\theta_{1-}$ where the $\pm$ denotes the chirality in the transverse space, i.e. under $SO(2)$. In terms of $\Gamma$ matrices, $\Gamma^3 \theta_{1+} = \theta_{1-}  $ and $\Gamma^{\bar{3}} \theta_{1-} = \theta_{1+} $. The four dimensional fermions can be obtained via dimensional reduction of $\theta_{1+}$ and $\theta_{1-}$, according to the cohomology classes of the cycle wrapped by the brane, as depicted below:
\begin{eqnarray}
\theta_{1+} &=& \displaystyle \sum_a \psi^a _{\pm \pm +} \otimes \chi^a _{\pm \pm +} \nonumber\\
\theta_{1-} &=&  \displaystyle \sum_a \psi^a _{\pm \pm -} \otimes \chi^a _{\pm \pm -},
\end{eqnarray}
where the $\psi^a$ are $4d$ spinors while the $\chi^a$ are internal spinors; the index $a$ simply counts the number of $4d$ spinors. The unspecified $\pm \pm$ indices correspond their chirality under $SU(2)$, i.e. corresponding to their behaviour under the action of $\Gamma^1$ and $\Gamma^2$. This allows us to group all the fields precisely as done in \cite{wrase1, wrase2}. We define 
\begin{eqnarray}
\lambda^0 &=& \sum \psi^a _{---} \;\;,\;\; \lambda^{\bar{0}} = \sum \psi^a _{+++} \nonumber\\
\lambda^1 &=& \sum \psi^a _{+--} \;\;,\;\; \lambda^{\bar{1}} = \sum \psi^a _{-++} \nonumber\\
\lambda^2 &=& \sum \psi^a _{-+-} \;\;,\;\; \lambda^{\bar{2}} = \sum \psi^a _{+-+} \nonumber\\
\lambda^3 &=& \sum \psi^a _{--+} \;\;,\;\; \lambda^{\bar{3}} = \sum \psi^a _{++-} . 
\end{eqnarray}
We can now perform the fermion decomposition exactly as in \cite{wrase1, wrase2}, except now the fermions $\lambda$ actually refer to the \emph{set} of fermions which transform according the corresponding chirality. We have
\begin{eqnarray} 
 \frac{\sqrt{2}}{12}\bar \theta^1 \Gamma^{MNP}\hat{\cal G}_{MNP}\theta^1&= &\bar \lambda_+^\zp\lambda_+^\zp \hat{\cal G}_{123}+\bar \lambda_-^{\zm}\lambda_-^{\zm}\hat{\cal G}_{\bar 1\bar 2\bar 3}
  +\left( \bar \lambda_+^{\zp}\lambda_+^{\ip}\hat{\cal G}_{ij\jb}-\bar\lambda_-^{\zm} \lambda_-^{\im}\hat{\cal G}_{\ib j \jb} \right)\delta^{j \jb}
\nonumber\\
&&+\ft12\left(\bar \lambda_+^{\ip}\lambda_+^{\jp}\varepsilon_{jk\ell}\hat{\cal G}_{ i \bar k\bar \ell}+\bar \lambda_-^{\im}\lambda_-^{\jm}\varepsilon_{\jb\bar k\bar \ell}\hat{\cal G}_{\ib k\ell}\right)\delta^{k\bar k}\delta^{\ell\bar \ell}\,,
 \label{gammaGdecomp}
\end{eqnarray}
where in our case $\hat{\cal G}_{MNP}\equiv \left(  \mathcal{G}_3 ^{\rm ISD2}  - \bar{\cal G}_3 ^{\rm ISD2} \right)_{MNP} $, and in an abuse of notation, we now use $M,N,P$ to refer to internal space, $M=4,5,...,9$.

The $\mathcal{G}_3$ flux must be (2, 1) and primitive, since we only want supersymmetry to be broken by the presence of the brane. This on its own immediately implies that $\lambda^0$ remains massless and that the mass cross-terms with $\lambda^i$ vanish as well, as in the $\overline{\rm D3}$ case. The additional feature that the flux which couples to the fermions is `ISD2' further reduces the allowed components to only those that have a $\bar{3}$ index, and hence the only non-vanishing mass terms are:
\begin{equation}
m_{3} = m_{ \bar{3} } \propto \left({\cal G}_3^{\rm ISD2}\right)_{12\bar{3}}, 
\end{equation}
where $\lambda^3$ gets its mass from ${\cal G}_3^{\rm ISD2}$ while $\lambda^{\bar{3}}$ gets its mass from $\bar{\cal G}_3^{\rm ISD2}$. The other fermions remain massless, i.e.
\begin{equation}
m_{0}=m_i=m_{0i}=m_{ij}=0 \;\;;\;\;i,j=1,2,
\end{equation}
and similarly for barred indices.

Thus the resulting four-dimensional massless fermionic field content consists of $\lambda^0$, $\lambda^1$ and $\lambda^2$. We emphasize that the $\lambda$'s refer to \emph{sets} of 4d fermions, the precise details of which can be found via dimensional reduction. Thus there are \emph{many} massless fermions in this case, in contrast to the $\overline{\rm D3}$ in a GKP background, which has only one \cite{wrase2}. However, both examples illustrate how supersymmetry is broken spontaneously by a probe anti-brane. Finally, we note that the bosonic field content on the brane can be taken care of as in the $\overline{\rm D3}$ case, by placing the $\overline{\rm D7}$ on an O7 plane.

\subsection{Inclusion of $\mathcal{F}$}
\label{sec:curlyF}

There is good reason to study non-zero $\mathcal{F}$: worldvolume fluxes on D7 branes generate D-terms and F-terms in the 4d theory  \cite{Jockers:2005zy}, and may even allow for de Sitter solutions along the lines of \cite{Burgess:2003ic}. With this in mind, let us see what happens on the anti-brane side of this story, i.e. what happens when we allow worldvolume fluxes on a $\overline{\rm D7}$.  Non-zero $\mathcal{F}$ modifies our previous analysis in two ways. One, it modifies the kinetic term via the matrix ${M}_{ab}$ 
defined earlier in \eqref{Mab} and two, it also modifies the $\kappa$-symmetry projector, which in turn induces new mass terms. 

The equations of motion require $\mathcal{F}$ to be anti self-dual on the cycle wrapped by the anti-brane, which we  take to be in the ($x^4, x^5, x^6, x^7$) directions, with $\epsilon_{4567}=-1$ to be consistent with our conventions in the previoius section. A judicious choice of vielbeins along the cycle can put the flux into the simple form,
\begin{equation}
\mathcal{F}=f(e_4\wedge e_5 + e_6 \wedge e_7).
\end{equation}
Note that in this approach we first choose a worldvolume flux, which then guides our choice of vielbeins and complex structure. This of course also affects the spacetime $\Gamma$-matrices and the definitions of the fermions in the $SU(3)$ triplet. At the end of the day, this amounts to an $SU(3)$ transformation and does not affect the number of massless fermions, which is what we are ultimately interested in, nor does it affect the masses of the massive ones.

The modified kinetic term can be recast as a canonical kinetic term plus a generalized electromagnetic coupling by a (generally non-isotropic) rescaling of the vielbeins, as described in \cite{martucci}. For our above choice of $\mathcal{F}$, the rescaling of the vielbeins to obtain a canonical kinetic term is simple. The matrix $M=g+\mathcal{F}$ now has off-diagonal terms, and in the vielbein basis its inverse is given by,
\begin{align}
M^{-1}=\frac{1}{1+f^2}\left(\begin{matrix} 1 &- f & 0 & 0\\ f & 1 & 0& 0 \\ 0 & 0 & 1 & -f \\ 0 & 0 & f & 1 \end{matrix}\right).
\end{align}
By defining rescaled vielbeins,
\begin{equation}
\hat{e}^m=\frac{1}{\sqrt{1+f^2}}e^a \qquad m=4,5,6,7,
\end{equation}
the kinetic term becomes
\begin{equation}
\bar{\theta}\Gamma_m \mathcal{D}_n M^{mn} \theta = \bar{\theta}\left( \hat{g}^{mn} + \hat{\mathcal{F}}^{mn} \right){\Gamma}_m {\mathcal{D}}_n \theta ,
\end{equation}
where the `hatted' quantities are expressed in terms of the rescaled vielbeins, e.g. $\hat{g}^{mn}=\eta^{jk}\hat{e}^m_j \hat{e}^n_k$.  We see that the kinetic term splits into a canonical kinetic term and a derivative coupling of the fermions to the worldvolume flux.

This derivative coupling complicates the dimensional reduction of $\theta_1$. The underlying SU(3) structure guarantees that there is are solutions to $g^{mn}\Gamma_m D_n \chi_6 = 0$, i.e. there exist zero-modes of the Dirac operator on the internal space, however it will generically \emph{not} be true that there are solutions to $(g^{mn} +{\cal F}^{mn})\Gamma_m D_n \chi_6 =0$, particularly for non-small ${\cal F}$. If no zero-modes exist for this `modified Dirac operator' then there will be no massless degrees of freedom. Thus the effect of the modified kinetic terms is to give mass to some, if not all, of the fermions.

We  still have yet to consider the modification of the couplings to $\mathcal{G}_3$. Before doing so, we must incorporate the rescaling of the vielbeins that we performed. This is simply done by putting a factor of $\sqrt{1+f^2}$ for every lower index along the brane directions in all the quantities. To avoid notation clutter, we will assume for the remainder of this section that the spacetime fluxes are implicitely `hatted' and contractions are made using the rescaled metric. This rescaling ultimately does not affect the tensor structure of the fluxes, and therefore will not affect which fermions acquire masses.

The inclusion of $\mathcal{F}$ also modifies the $\kappa$-symmetry projector, in the following way:
\begin{align}
\Gamma_{\overline{\rm D7}}&=\frac{1}{\sqrt{|g+\mathcal{F}|}}\left[-i\sigma_2\Gamma_{01234567}+\sigma_3 i \sigma_2 (\Gamma_{012345}\mathcal{F}_{67}-\Gamma_{012367}\mathcal{F}_{45})-i\sigma_2 \Gamma_{0123}\mathcal{F}_{45}\mathcal{F}_{67}\right] \nonumber \\
&=\frac{1}{\sqrt{|g+\mathcal{F}|}}\left[-i\sigma_2\Gamma_{01234567}+\hat{f}\sigma_3 i \sigma_2 (\Gamma_{012345}-\Gamma_{012367})-i\sigma_2 \Gamma_{0123}\hat{f}^2\right],
\end{align}
which in turn modifes the relation between $\theta_{1,2}$ imposed by the gauge fixing condition $\bar{\theta}(1+\Gamma_{\overline{\rm D7}})=0$, in the following way:
\begin{equation}
\theta_2=\left[\Gamma_{01234567}+\hat{f}(\Gamma_{012345}-\Gamma_{012367})+\hat{f^2}\Gamma_{0123}\right] \theta_1.
\end{equation}
The outcome of all these changes is that now new coupling arise as:
\bg\label{newcoup} 
&&\bar{\theta}_1 e^{-\phi} \Big[\Gamma^{mna} \left({\cal F}_{mna}\Gamma_{0...7}+ e^{-\phi}{\cal H}_{mna}\right)+\hat{f} \Gamma^{mab}\left({\cal F}_{mab}(\Gamma_{012345}-\Gamma_{012367})+e^{-\phi}{\cal H}_{mab}\right)\nonumber\\ 
&&~~~~~~~~~~~~~ + \hat{f}^2\Gamma^{mnl}\left({\cal F}_{mnl}\Gamma_{0123}+e^{-\phi}{\cal H}_{mnl}\right) \Big]\theta_1,
\nd
where the indices ($m,n,l$) now take values 4,5,6,7 and ($a,b$) as before take values (8,9).

These new terms include fluxes that have one leg or all three legs along the brane, which were not presence for $\mathcal{F}=0$. In fact, the last term is the coupling we get for an $\overline{\rm D3}$ brane. This is to be expected, since worldvolume fluxes induce lower-dimensional brane charge. The term linear in $\hat{f}$ is the coupling due to the induced five-brane charge and is similar to what we would obtain if we studied an $\overline{\rm D5}$ in a GKP background. It produces couplings to fluxes which obey a self-duality condition in the directions transverse to the cycles threaded by the flux. As in the pure $\overline{\rm D7}$ case, this simply restricts which subset of fermions get masses and produces no new unexpected couplings. The presence of the $\overline{\rm D3}$-like coupling means that the $SU(3)$ triplet fermions will generically all acquire a mass (in addition to any mass they receive from the modified kinetic term), though some may remain massless due to the specific form of the flux as we saw in the previous section. The singlet fermions, however, receive no new ${\cal G}_3$ induced mass, for the same reason as before: its mass term does not arise from primitive (2, 1) fluxes, which we require by construction. However, as mentioned already, the singlet \emph{does} in general receive a mass from the modified kinetic term, and hence there will generically remain no massless degrees of freedom.

\subsection{Effect of more general ${\cal F}_5$}
\label{sec:F5}
 
Before we close this section we wish to comment on how the scenario changes once we allow for more general ${\cal F}_5$. The combination 
\bg\label{tagra}
\widetilde{\cal F}_5 = {\cal F}_5+ {\cal B}_2\wedge {\cal F}_3 + {\cal C}_2 \wedge {\cal H}_3, \nd 
needs to be self-dual in the full $10d$ space. If we demand that the 3-form fluxes 
have {only} {\it one} leg transverse to the brane, which is necessary for them to give fermion masses, then the 5-form flux must have a leg off the brane as well and therefore will not generate a mass for the fermions! Conversely, if ${\cal F}_5$ is entirely along the brane directions, the corresponding 3-forms will not be of the appropriate form to generate masses. It is therefore possible to consider embeddings of the $\overline{\rm D7}$ such that only one or the other type of mass contributions are present or combine both. 

Let's consider a non-zero ${\cal F}_{0123m}$ component, where $m$ is along the brane worldvolume. The fermion decomposition analysis is very similar to before. The contribution to the action is of the form

\begin{eqnarray}
 \frac{\sqrt{2}}{12}\bar \theta^1 \Gamma^{MNP}\varepsilon_{MNPQ}{\cal F}_{0123}^{\quad \; \;Q}\theta^1&= &\left( \bar \lambda_+^{\zp}\lambda_+^{\ip}\varepsilon_{ij\jb \bar{k}}{\cal F}_{0123}^{\quad \; \; \bar{k}}-\bar\lambda_-^{\zm} \lambda_-^{\im}\varepsilon_{\bar{i}j \bar{j} k}{\cal F}_{0123}^{\quad \; \; k} \right)\delta^{j \jb}\\
&&+\ft12\left(\bar \lambda_+^{\ip}\lambda_+^{a}\varepsilon_{a k\ell}\varepsilon_{i \bar{k} \bar{\ell} m}{\cal F}_{0123}^{\quad \; \; m}+\bar \lambda_-^{\im}\lambda_-^{\bar{a}}\varepsilon_{\bar{a} \bar k\bar \ell}\varepsilon_{\bar{i} k \ell \bar{m}}{\cal F}_{0123}^{\quad \; \; \bar{m}}\right)\delta^{k\bar k}\delta^{\ell\bar \ell}, \nonumber
\end{eqnarray}
where the $i,j,k...$ indices are restricted to lie along the brane (but $a$ has no such restriction). This results in non-zero $m_{13}, m_{23}$ and even more notably $m_{01}, m_{02}$. Note that $m_{11}, m_{12}, m_{22}$ remain vanishing, so even when both the 3-form and the 5-form fluxes contribute mass terms, there is still a massless degree of freedom remaining.

Finally, the modification of the $\kappa$-projector in the scenario with worldvolume fluxes does not introduce new contributions from the 5-form flux. Indeed, the second term in $\Gamma_{\overline{\rm D7}}$, which gives the coupling to the induced five-brane charge, can only conspire to give 3 or 7 gamma matrices inside the resulting fermion bilinear if ${\cal F}_5$ has two legs in the internal space, but it must have four legs along the spacetime directions. Similarly, the third term necessarily results in a single gamma matrix, yielding a vanishing bilinear, exactly as in the $\overline{\rm D3}$ case. Note however, that in combining both worldvolume fluxes and an ${\cal F}_5$ without transverse legs results in all the fermions acquiring a mass.

Let us also note that if we had taken the internal space to be a non-K\"ahler resolved conifold with fractional branes, and then inserted a 
${\overline{\rm D7}}$-brane wrapping a four-cycle inside the non-K\"ahler space, the background fluxes and also the physics would have been quite different. We will however not 
explore this further here, but instead go to another interesting aspects of our analysis: the all-order fermionic action on a ${\overline{\rm D3}}$-brane.

\section{Towards the $\kappa$-symmetric All-Order Fermionic Action for a $\overline{\rm D3}$-brane}

The previous two sections detailed the spontaneous breaking of supersymmetry by probe anti-branes in otherwise supersymmetric compactifications. The starting point of both of these analyses has been the fermionic brane action at lowest order in $\theta$, which takes a manifestly $\kappa$-symmetric form.

We would now like to see if this result continues to hold at higher orders in $\theta$. As we will see, the answer to this question is in affirmative, 
and to show this we need only minimal knowledge of brane actions\footnote{See \cite{Komsei}, \cite{kuzen}, and \cite{Bandos:2015xnf}, for more recent related works on the Volkov-Akulov actions.}. 
In particular, we can use string dualities to deduce the structure of the all order fermionic action, without needing precise information as to the form of the higher order operators. To do so, we will define a (completely general) fermionic completion of the $\overline{\rm D3}$-brane action, as was done at lowest order in $\theta$ in \cite{marolf1, marolf2}, and use certain duality tricks to generate the higher order fermionic counterparts of the bosonic fields. Note that under RG the higher order terms are generically irrelevant, but they are
nevertheless needed to realize the full $\kappa$-symmetry.

The bosonic components 
of the NS and RR sectors are connected by the type IIB equations of motion, and therefore once a certain set of field components are known, others can be generated 
from the corresponding EOMs. On the other hand, for the fermionic components no additional work is needed: knowing the fermionic fields ($\theta,   
\bar{\theta}$) and the bosonic fields, one should be able to predict the fermionic completions of the bosonic fields to all orders in $\theta$ and $\bar{\theta}$.
This means the fermionic completions of higher $p$-form fields 
should {\it at least} be related to the lower $p$-form field (including the graviton, anti-symmetric tensor and dilaton) 
by certain U-duality transformations at the self-dual points $g_s = 1$ and $R_i = 1$ for $i = 1,.., 2k$ with $R_i$ being the radii of the compact directions. 
To see why this is the case, let us study two corners of type IIB moduli space.

\vskip.05in

\begin{figure}[htb]
        \begin{center}
\includegraphics[height=9cm]{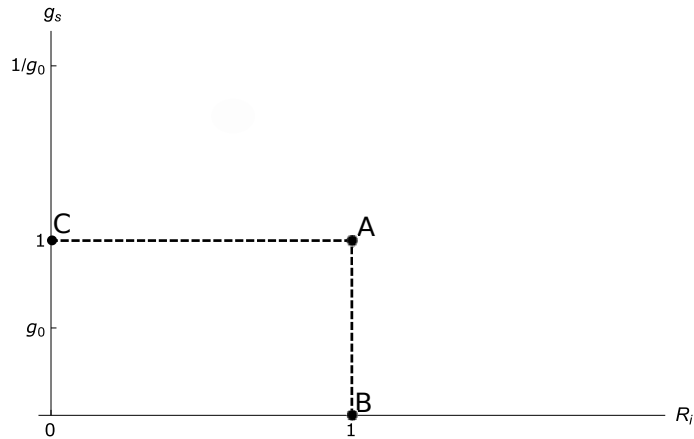}
        \caption{The Type IIB moduli space with the self-dual point denoted by $A$. The point $B$ is for all $R_i = 1$ and the point $C$ is for $g_s = 1$. Our duality mappings are 
defined for the point $A$. Going away from the point $A$ in any direction in the moduli space will imply switching on non-trivial values for the axio-dilaton.}
\label{IIBmod}        
\end{center}
        \end{figure}

\noindent $\bullet$ We can go to $g_s = e^\phi = 1$ point where we should be able to exchange ${\bf B}^{(1)}_{mn}$ with ${\bf B}^{(2)}_{mn}$, as shown as point $C$ in 
{\bf figure \ref{IIBmod}}.

\vskip.05in

\noindent $\bullet$ We can go to self-dual radii of the compact target space $R_i = 1$ where we should be able to exchange the $p$-form fields with ($p + 2k$)-form fields,
as shown as point $B$ in {\bf figure \ref{IIBmod}}.

\vskip.1in 

\noindent This is only possible if at least a subset of the  fermionic counterparts of the ($p + 2k$)-form fields are the ones got 
via U-duality transformations. This trick could then be used to generate all the fermionic counterparts of the higher form fields at least at the self-dual corner
$g_s = R_i = 1$ of type IIB moduli space, i.e around the point $A$ in {\bf figure \ref{IIBmod}}. Once we move away from the self-dual point, we can study the fermionic 
counterparts of the bososnic fields at generic point in the type IIB moduli space.   

On the other hand the scenario is subtle in the presence of branes. It is known that the D3 or the ${\overline {\rm D3}}$-branes are S-duality neutral although the world-volume 
degrees of freedom differ. However they are not T-duality neutral. The other D-branes (or NS-branes) are neither S nor T-duality neutrals. So, to effectively use the 
duality trick, no branes should be present. This is good because now we can determine the fermionic completion of the background without worrying about the backreactions from the branes, 
and then 
insert D-branes to study the world-volume theory. 

\subsection{Towards all-order $\theta$ expansion from dualities}

Let us now proceed with our analysis. We start by redefining the all order fermionic completion of type IIB scalar fields in the following way:
\bg\label{bosfields} 
{\bf \Phi}^{(i)} &=& \varphi^{(i)} + \bar{\theta} \Delta^{(i)} \theta \\ 
&\equiv& \varphi^{(i)} + \displaystyle \sum_j \prod _{k=1} ^{j}\bar{\theta} \Delta^{(i)jk} \theta \nonumber\\
&= & \varphi^{(i)} + \bar{\theta} \Delta^{(11i)}\theta + \bar{\theta} \Delta^{(21i)}_{m..p} \theta \bar{\theta}\Delta^{(22i)}_{q..n}\theta g^{pq}.. g^{mn} 
+ \bar{\theta} \Delta^{(31i)}_{m..p} \theta \bar{\theta}\Delta^{(32i)}_{q..l}\theta \bar{\theta} \Delta^{(33i)}_{s..n}\theta g^{pq}.. g^{ls}.. g^{mn} + ...
\nonumber \nd
where ${\bf \Phi}^{(1)} = \phi_B$ and ${\bf \Phi}^{(2)} = C^{(0)}$ 
are the dilaton and the axion respectively; and the dotted terms are of ${\cal O}(\theta^{8})$. The
fermion products in \eqref{bosfields} are defined in terms of components in the following way:
\bg\label{ferbilin}
\bar{\theta} \Delta^{(21i)}_{m..p} \theta \bar{\theta}\Delta^{(22i)}_{q..n}\theta \equiv
\bar{\theta}^\alpha \Delta^{(21i)}_{m..p\alpha\beta} \theta^\beta \bar{\theta}^\delta\Delta^{(22i)}_{q..n\delta\gamma}\theta^\gamma, \nd
where the Greek indices span the 32 (complex) component\footnote{Or 64 real component. Note that the series in \eqref{bosfields} and in the following, 
terminate at some finite number of terms because 
of finite number of fermionic components as well as because of the  Grassmannian nature of the fermions.} 
fermions $\theta$. The IIB spinor $\theta$ is a doublet of 16 (complex) component Majorana spinors 
of the same chirality, i.e this doublet is a 32 component Majorana spinor but is \emph{not} Weyl.  We decompose $\theta$ into the two 16 (complex) component fermions $\theta_1$ and 
$\theta_2$ as:
\bg\label{theta}
\theta = \left(\begin{matrix}\theta_1\\ \theta_2 \end{matrix}\right), \nd 
with $\theta_2$ generically non-vanishing. 
The $\Delta^{(ab i)}$ are all operators that can be represented in the matrix form in the following way:
\bg\label{deltai}
\Delta^{(i)} \equiv \left(\begin{matrix} \Delta^{(11i)} & \Delta^{(12i)} & \Delta^{(13i)} & ....\\ \Delta^{(21i)} & \Delta^{(22i)} & \Delta^{(23i)} & ....\\
\Delta^{(31i)} & \Delta^{(32i)} & \Delta^{(33i)} & ....\\
.... & {} & {} & {} \end{matrix}\right), \nd
where every element of the matrix should be viewed as an operator with its own matrix representation in some appropriate Hilbert space. The complete form of the 
matrix \eqref{deltai} is not known, but a few elements have been worked out in the literature \cite{marolf1, marolf2, bergkallosh, martucci}. For example it is known
that:
\bg\label{known} 
\Delta^{(111)}\theta = -{i\over 2} \bar{\delta}\lambda~\theta, ~~~~~~~ \Delta^{(112)}\theta = {1\over 2} e^{-\phi}\sigma_2\bar{\delta}\lambda~\theta, \nd
where $\bar{\delta}\lambda$ is the supersymmetric variation of the type IIB spinor $\lambda$ in the presence of an $\overline{\rm D3}$
and $\sigma_2$ is the second Pauli matrix that act on the $\theta_{1, 2}$
components of \eqref{theta}. 
It should also be clear, from
the way we constructed the matrix, that:
\bg\label{clear} \Delta^{(ab i)} ~ = ~ \Delta^{(ba i)}. \nd
Additionally, in the ensuing analysis we will resort to the following simplification: instead of considering the $\Delta^{(abi)}$ operators to have an arbitrary rank $q$ as 
$\Delta^{(abi)}_{m_1 m_2....m_q}$ for $a\ge 2$, we will only take them to have a maximal rank $2$. As will be clear from the context, this simplification will not change any of the physics, 
and one may easily switch to arbitrary rank $\Delta^{(abi)}$ operators without loss of generalities. On the other hand, this simplification will avoid unnecessary cluttering of indices. 
Henceforth unless mentioned otherwise, we will take only this simplified version. 

With this in mind, 
let us now consider the type IIB metric $g_{mn}$. We can expand the all order fermionic completion in a way analogous to the scalar field:
\bg\label{meturi}
{\bf G}_{mn} &=& g_{mn} + \bar{\theta} M_{(mn)} \theta \\
& = & g_{mn} + \bar{\theta} M^{(11)}_{(mn)} \theta + g^{pq} \bar{\theta} M^{(21)}_{(m \vert p} \theta \bar{\theta} M^{(22)}_{q \vert n)} \theta 
+ g^{pq} g^{ls} \bar{\theta} M^{(31)}_{(m \vert p} \theta \bar{\theta} M^{(32)}_{ql} \theta \bar{\theta} M^{(33)}_{s \vert n)} \theta 
+ {\cal O}(\theta^{8}) \nonumber, \nd 
which is again a sum over products of contractions of the fermions with matrix elements of the operator $M_{mn}$. The four-component $M$ operator can be written using two bosonic and two fermionic components as:
\bg\label{mmn}
M_{(mn)\alpha\beta} = M^{(11)}_{(mn)\alpha \beta} + M^{(21)p}_{(m \vert \alpha \gamma} \theta^\gamma \bar{\theta}^\delta M^{(22)}_{p \vert n)\delta\beta} + 
M^{(31)p}_{(m \vert \alpha \gamma} \theta^\gamma \bar{\theta}^\delta M^{(32)}_{ps\delta\sigma} \theta^\sigma \bar{\theta}^\rho M^{(33)s}_{n)\rho\beta} + ...\nd
where the first term in the above expansion is well-known in terms of the supersymmetric variation of Rarita-Schwinger 
fermion $\psi_m$ \cite{marolf1, marolf2, bergkallosh, martucci}:
\bg\label{1terma}
M^{(11)}_{\alpha\beta (mn)}~ = ~ - i \Gamma_{\alpha\beta (m} D_{n)} ~ \equiv ~ -i \Gamma_{\alpha\beta (m} \bar{\delta} \psi_{n)}. \nd 
The anti-symmetric rank two tensor can also be expanded in terms of the fermionic components like the symmetric tensor \eqref{meturi}. We can define
${\bf B}_{mn}^{(i)}$ as the generalized anti-symmetric tensors, where ${B}_{mn}^{(1)} = B_{mn}$ and ${B}_{mn}^{(2)} = C^{(2)}_{mn}$ as the NS and RR two-forms 
respectively, using certain anti-symmetric tensor $N^{(i)}_{[mn]}$ in the following way:
\bg\label{bumun}
{\bf B}^{(i)}_{mn} &=& B^{(i)}_{mn} + \bar{\theta} N^{(i)}_{[mn]} \theta \\
& = & B^{(i)}_{mn} + \bar{\theta} N^{(11i)}_{[mn]} \theta + g^{pq} \bar{\theta} N^{(21i)}_{[m|p} \theta \bar{\theta} N^{(22i)}_{q| n]} \theta + g^{pq} g^{ls} \bar{\theta} N^{(31i)}_{[m | p} \theta \bar{\theta} N^{(32i)}_{ql} \theta \bar{\theta} N^{(33i)}_{s| n]} \theta 
+ {\cal O}(\theta^{8}) . \nonumber \nd  
To see the connection between $M_{(mn)}$ and $N^{(i)}_{[mn]}$ operators let us revisit the T-duality rules of \cite{BHO, fawad}. The powerful thing about the 
fermionic completion is that the T-duality rules follow {\it exactly} the formula laid out for the bosonic fields, except now all the fields are replaced
by their fermionic completions. This can be illustrated as\footnote{There seems to be two ways of analyzing the T-duality transformations in the literature. 
One, is to assume that the Buscher's rules are 
{\it exact} to all orders in $\alpha'$ and only the supergravity fields receive $\alpha'$ corrections. This way, the Busher's rule could be used to study supergravity field transformations
order by order in $\alpha'$. Two, both the T-duality transformations {\it and} the supergravity fields receive $\alpha'$ corrections. There is some confusion of which one should be 
considered, but in our opinion the more conservative picture is the latter one where {\it both}, the T-duality rules as well as the supergravity fields, receive $\alpha'$ corrections. Since 
T-duality transformations preserve supersymmetry, the $\alpha'$ corrections to the T-duality transformations would imply $\alpha'$ corrections to the supersymmetry transformations: a result
consistent with the known facts. See for example \cite{bergkal, berg} for the lowest order corrections, where somewhat similar arguments have appeared;
and \cite{olaf} for more recent discussions. However as we will see soon, our results will not be very sensitive to this.}:
\bg\label{trule}
&&\widetilde{\Phi}^{(1)} = \Phi^{(1)} -\frac 12 {\rm ln}\; {\bf G}_{xx}\quad\quad\quad\quad\quad\qquad\quad\quad\quad\hspace{.13cm} 
\widetilde{{\bf G}}_{xx} = {1\over {\bf G}_{xx}}
\nonumber\\
&&\widetilde{{\bf G}}_{ m n} = {\bf G}_{ m n}
- { {\bf G}_{m x} {\bf G}_{ n x}
- {\bf B}^{(1)}_{ m x} {\bf B}^{(1)}_{n x}
\over {\bf G}_{xx}}
\qquad\qquad 
\widetilde{{\bf G}}_{ m x} ={ {\bf B}^{(1)}_{m x}
\over {\bf G}_{xx}}\\
&&\widetilde{\bf B}^{(1)}_{m n}={\bf B}^{(1)}_{ m n}
-{{\bf B}^{(1)}_{m x} {\bf G}_{ n x}-{\bf G}_{m x}
{\bf B}^{(1)}_{n x}\over {\bf G}_{xx}} \quad\qquad\quad
\widetilde{\bf B}^{(1)}_{m x} ={ {\bf G}_{m x} \over {\bf G}_{xx}} \nonumber 
\nd
where $x$ is the T-duality direction. From the T-duality rule we see that, in the presence of cross-terms of ${\bf G}$ in type IIA, 
${\bf B}^{(1)}$ could be generated in type IIB using \eqref{trule}. Since both IIA or IIB metric uses $M_{(mn)}$, this is possible if:
\bg\label{NorM}
\bar{\theta} N^{(1)}_{[mx]} \theta ~ \equiv \bar{\theta} ~c \sigma^p_3 M_{[mx]}\theta, \nd
where the operator $M_{[mx]}$ is now expressed with respect to the T-dual fields, i.e the IIB bosonic fields. 
We have also inserted the third Pauli matrix $\sigma_3$ in \eqref{NorM}, with $p = 1$ or 2, to take care of certain subtleties that will be explained 
later\footnote{See discussions after \eqref{wayout}.}, 
and $c$ is a constant matrix. The only constant matrices for our case, that do not change the chirality, are the identity and
the chirality matrix $\Gamma^{10}$, so we will choose $c = \Gamma^{10}$. Since we can make T-duality 
along any direction, $x$ appearing in \eqref{NorM} could span all directions. This means we can generalize \eqref{NorM} to the following:
\bg\label{NM}
\bar{\theta} N^{(1)}_{[mn]} \theta ~ \equiv \bar{\theta} ~\sigma^p_3 \otimes \Gamma^{10}  M_{[mn]}\theta, \nd
implying that the symmetric matrix $M_{(mn)}$ determines the generalized metric ${\bf G}_{mn}$, whereas the anti-symmetric matrix $M_{[mn]}$ determines the 
generalized B-field ${\bf B}^{(1)}_{mn}$.  In terms of components, we expect:
\bg\label{jhor}
N^{(111)}_{\alpha\beta [mn]} &= & - i \sigma_3 \otimes \Gamma^{10} \Gamma_{\alpha \beta[m} \delta\psi_{n]}, \nd
which is consistent with the results in \cite{marolf1, marolf2, bergkallosh, martucci}. However the relation \eqref{NM} predicts the form of \emph{all} the operators appearing 
in \eqref{bumun} once all the corresponding operators appearing in \eqref{meturi} are known, not just the component given above.  

To find the form of ${\bf B}^{(2)}_{mn}$, or the operator $N^{(2)}_{[nm]}$, we will use the T-duality trick discussed above, assuming that the T-duality rules go
for the RR fields with fermionic completions exactly as their bosonic counterparts \cite{marolf1, marolf2}. To proceed we will need ${\bf \Phi}^{(2)}$ and 
${\bf B}_{mn}^{(1)}$ from \eqref{bosfields} and \eqref{bumun}, rewritten as:
\bg\label{joba}
{\bf \Phi}^{(2)} = C^{(0)} + \bar{\theta} \sigma_2 \widetilde{\Delta}^{(2)} \theta, ~~~~~~
{\bf B}_{mn}^{(1)} = B_{mn} + \bar{\theta} \sigma_3 \otimes \Gamma^{10}{M}_{[mn]} \theta, \nd
where we have extracted a Pauli matrix $\sigma_2$ in defining $\Delta^{(2)} = \sigma_2 \widetilde{\Delta}^{(2)}$. The other components appearing in \eqref{joba} are
the corresponding bosonic backgrounds. The T-duality rules for the RR fields are given as\footnote{As before, we expect the T-duality rules for the RR fields to also receive $\alpha'$ 
corrections. We will discuss the consequence of this on our analysis soon.}:
\bg\label{RRfields}
&& \widetilde{ {\bf C}}^{(n)}_{x m_2\cdots m_n}={\bf C}^{(n-1)}_{ m_2\cdots m_n}-(n-1)\widetilde{{\bf B}}^{(1)}_{x[m_2}{\bf
C}^{(n-1)}_{|x|m_3\cdots m_n]}, \nonumber\\
&& \widetilde{{\bf C}}^{(n)}_{ m_1\cdots m_n}={\bf C}^{(n+1)}_{x
  m_1\cdots m_n}-n{\bf B}^{(1)}_{x[m_1}\widetilde {\bf C}^{(n)}_{|x|m_2\cdots  m_n]}.
\nd
There are now two possible ways to get the fermionic part of ${\bf B}_{mn}^{(2)}$: we can T-dualize twice the scalar ${\bf \Phi}^{(2)}$ using the T-duality 
rule \eqref{RRfields}, and we can S-dualize ${\bf B}^{(1)}_{mn}$. Let us start by discussing the first possibility, namely 
the T-duality way of getting part of ${\bf B}_{mn}^{(2)}$. 
T-dualizing once we get a vector field in type IIA as:
\bg\label{vect}
\widetilde{\bf C}^{(1)}_x = {\bf \Phi}^{(2)}, \nd
and then another T-duality will give us the required RR two-form field in the following way:
\bg\label{rere}
\hat{\bf C}^{(2)}_{yx} = \widetilde{\bf C}^{(1)}_x - \hat{\bf B}^{(1)}_{yx} \widetilde{\bf C}^{(1)}_y = {\bf \Phi}^{(2)} 
- \hat{\bf B}^{(1)}_{yx} \widetilde{\bf C}^{(1)}_y = {\bf \Phi}^{(2)} \nd
because $\widetilde{\bf C}^{(1)}_y = 0$ according to \eqref{vect}, and therefore the field ${\bf \Phi}^{(2)}$ should determine the required two-form. However
before proceeding we should determine how the 32 component Majorana fermion \eqref{theta} change under the two T-dualities. It is easy to show that:
\bg\label{thetachange}
\theta ~ \to ~ \Sigma_1~\theta, ~~~~~~~~ \bar{\theta} ~ \to ~ \bar{\theta} ~\Sigma_2, \nd
where $\Sigma_i$ are two $32 \times 32$ component matrices, i.e. the act on the doublet basis, given in terms of the sixteen component Gamma 
matrices\footnote{We are using the flat-space $\Gamma$ matrices.} $\Gamma_x$ and $\Gamma_y$ by:
\bg\label{sigma12}
\Sigma_1 ~ = ~ \left(\begin{matrix} {\bf I}_{16} &{} & 0 \\ 0 &{} & ~~~~ \Gamma_x \Gamma_y \end{matrix}\right), ~~~~~~~~
\Sigma_2 ~ = ~ \left(\begin{matrix} {\bf I}_{16} & {} & 0 \\ 0 & {} & ~~~~ \Gamma_y \Gamma_x \end{matrix}\right),\nd 
and leading to the following set of algebras that will be useful soon:
\bg\label{algebras}
&&\Sigma_2 (\sigma_2 \otimes {\bf I}_{16}) \Sigma_1 ~ = ~ \sigma_3 \sigma_2 \otimes \Gamma_x \Gamma_y, ~~~~~~~~ \Sigma_2 \cdot \Sigma_1 ~ = ~ {\bf I}_{32} \nonumber\\
&& \Sigma_2 \left(\begin{matrix} 0 &{}& \mp i{\bf C} \\ \pm i{\bf C} & {} & 0\end{matrix}\right) \Sigma_1 = 
\left(\begin{matrix} \pm {\bf C} & {} & 0 \\ 0 & {} & \pm {\bf C} \end{matrix}\right) (\sigma_3 \sigma_2 \otimes \Gamma_x \Gamma_y) \nonumber\\
&& \Sigma_2 \left(\begin{matrix} \pm {\bf C} & {} & 0 \\ 0 & {} & \pm {\bf C} \end{matrix}\right) \Sigma_1 
= \left(\begin{matrix} \pm {\bf C} & {} & 0 \\ 0 & {} & \pm {\bf C}\end{matrix} \right), ~~~~~ \left(\sigma_3 \sigma_2\right)^2 = -{\bf I}_2 \nonumber\\
&& \left(\begin{matrix} {\bf 1} &{} & 0 \\ 0 &{} & ~~~~ \Gamma_b \Gamma_a \end{matrix}\right) (\sigma_3 \sigma_2 \otimes \Gamma_x \Gamma_y)
\left(\begin{matrix} {\bf 1} &{} & 0 \\ 0 &{} & ~~~~ \Gamma_a \Gamma_b \end{matrix}\right) = \sigma_2 \otimes \Gamma_x \Gamma_y \Gamma_a \Gamma_b .\nd 
Therefore using \eqref{rere} and \eqref{thetachange} with the algebras \eqref{algebras}, we can get one part of the two-form ${\bf B}^{(2)}_{mn}$ in the
following way:
\bg\label{chat}
\hat{\bf C}^{(2)}_{mn} = \bar{\theta} c \sigma_3 \sigma_2 \otimes \Gamma_{mn} \widetilde{\Delta}^{(2)} \theta, \nd
where, as before, we can take $c = \Gamma^{10}$ i.e the chirality matrix, and $\widetilde{\Delta}^{(2)}$ can either be expressed in terms of the T-dual
fields or the original fields. 

In deriving \eqref{chat} we haven't actually looked at the form of $\widetilde{\Delta}^{(2)}$. Depending on the representation
 of Gamma matrices in the definition of
$\widetilde{\Delta}^{(2)}$, our simple expression \eqref{chat} could in principle change to a more involved one. The scenario is subtle so let us tread carefully
here. We start by rewriting the RR scalar \eqref{bosfields} field as:
\bg\label{rrfield} 
{\bf \Phi}^{(2)} & = & C^{(0)} + \left(\bar{\theta} \sigma_2\right)^\alpha \widetilde{\Delta}^{(2)}_{\alpha\beta} \theta^\beta \nonumber\\
& = & C^{(0)} + (\bar{\theta} \sigma_2)^\alpha \widetilde{\Delta}_{\alpha\beta}^{(112)}\theta^\beta 
+ (\bar{\theta}\sigma_2)^\alpha \widetilde{\Delta}_{\alpha\chi m}^{(212)p} \theta^\chi 
\bar{\theta}^\sigma\widetilde{\Delta}_{\sigma\beta p}^{(222)m}\theta^\beta \nonumber\\
&& ~~~~~ + (\bar{\theta}\sigma_2)^\alpha\widetilde{\Delta}_{\alpha\gamma m}^{(312)p} 
\theta^\gamma 
\bar{\theta}^\sigma\widetilde{\Delta}_{\sigma\chi p}^{(322)l}\theta^\chi \bar{\theta}^\delta\widetilde{\Delta}_{\delta\beta l}^{(332)m}
\theta^\beta + {\cal O}(\theta^8), 
\nd
where we have assumed that the generic operator $\widetilde{\Delta}^{(ab2)m}_{\alpha\beta n}$ 
is constructed from the products of 16 dimensional Gamma matrices, the 
type IIB bosonic fields and covariant derivatives ${\bf A}_{16 \times 16}$ as:
\bg\label{onerep}
\left(\bar{\theta}\sigma_2^p\right)_\alpha \widetilde{\Delta}^{(ab2)}_{mn\alpha\beta}\theta_\beta \equiv 
\bar{\theta}\sigma_2^p \left(\begin{matrix} {\bf A}^{(ab)}_{16 \times 16} & {} & 0 \\ 0 & {} & {\bf A}^{(ab)}_{16 \times 16}\end{matrix}\right)_{mn} \theta, 
\nd 
where $p$ can be 0 or 1 depending on what fermion combination we are looking at in \eqref{rrfield}. Using our T-duality ideas, and using the Gamma matrix
algebras \eqref{algebras}, it is easy to see that the two-form \eqref{chat} appears naturally with an overall $\Gamma_m \Gamma_n$ matrix provided we impose:
\bg\label{acomm} [{\bf A}, ~ \Gamma_x \Gamma_y] = 0, \nd 
without loss of generalities as transformations with even number of Gamma matrices will not change any results. The puzzle however
is if \eqref{onerep} takes the following form:
\bg\label{tworep} 
\left(\bar{\theta}\sigma_2^p\right)^\alpha \widetilde{\Delta}^{(ab2)}_{mn\alpha\beta}\theta^\beta \equiv 
\bar{\theta}(\sigma_2^p \otimes {\bf I}_{16})\left( {\bf I}_2 \otimes {\bf A}^{(ab)} + \sigma_1 \otimes {\bf C}^{(ab)}  \right)_{mn} \theta 
= \bar{\theta}(\sigma_2^p \otimes {\bf I}_{16}) \left(\begin{matrix} {\bf A}^{(ab)}_{16 \times 16} & {} &  {\bf C}^{(ab)}_{16 \times 16} \\  
{\bf C}^{(ab)}_{16 \times 16} & {} & 
{\bf A}^{(ab)}_{16 \times 16}\end{matrix}\right)_{mn} \theta, \nonumber\\ \nd
where ($\sigma_1, {\bf I}_2$) are the first Pauli matrix and 2 dimensional identity matrix respectively; and 
${\bf C}_{16 \times 16}$ is another 16 dimensional matrix constructed out of Gamma matrices, IIB fields and covariant derivatives. 

To understand the consequence of the above mentioned representations of the operators, let us discuss a few additional Gamma matrix algebras under our T-duality
transformations:
\bg\label{gamre}
&& \Sigma_2 (\sigma_2 \sigma_1 \otimes {\bf C}) \Sigma_1 ~ =  -i \sigma_3 \otimes {\bf C} \nonumber\\
&& \Sigma_2(\sigma_1 \otimes {\bf C}) \Sigma_1 ~ = i (\sigma_2 \otimes {\bf C}\Gamma_x \Gamma_y) \nonumber\\
&&\Sigma_2 (\sigma_2 \otimes {\bf I}_{16}) ({\bf I}_2 \otimes {\bf A}) \Sigma_1~ =  ~ \sigma_3 \sigma_2 \otimes {\bf A}\Gamma_x \Gamma_y. \nd
Using these algebras, it is now easy to see that under T-dualities the operators \eqref{onerep} and \eqref{tworep} transform in the following way:
\bg\label{optransf}
&& \bar{\theta} \left({\bf I}_2 \otimes {\bf A} + \sigma_1 \otimes {\bf C} \right)\theta ~\to ~ \bar{\theta}\left({\bf I}_2 \otimes {\bf A} + i \sigma_2 \otimes {\bf C}  \Gamma_x \Gamma_y\right)\theta \nonumber\\
&& \bar{\theta} \sigma_2 \left({\bf I}_2 \otimes {\bf A} + \sigma_1 \otimes {\bf C}\right) \theta ~ \to ~ \bar{\theta}\left(\sigma_3 \sigma_2 \otimes {\bf A} \Gamma_x \Gamma_y - i  \sigma_3 \otimes{\bf C} \right) \theta, \nd
{}from where we see that the first terms in \eqref{optransf} are clearly 
consistent with the duality rules that lead us to the result \eqref{chat}. However it is the 
second term in the two expressions above in \eqref{optransf} which would {\it not} fit with the generic result \eqref{chat}. Clearly when ${\bf C} = 0$ this
problem does not arise.  

A way out of this conundrum is in fact clear from the transformations themselves. Existence of ${\bf C}_{16 \times 16}$  
in \eqref{tworep} would imply that this piece is T-duality 
neutral, and {\it doesn't} transform as a rank 2 tensor under T-duality. Thus this piece cannot be part of a RR axionic scalar whose T-duality transformations are well known. In fact
its neutrality to the T-duality transformation hints that ${\bf C}_{16 \times 16}$ 
could be a part of the NS scalar i.e the dilaton, unless of course    
we can use $\bar{\psi} \equiv \bar{\theta} \sigma_1$ 
to transform
\bg\label{wayout}
\bar{\psi} {\bf C} \otimes \sigma_1 \theta ~ \to ~ \bar{\theta} {\bf C} \theta, \nd
under two T-dualities. This way the issues raised in \eqref{optransf} will not arise and the generic result \eqref{chat} will continue to hold to arbitrary 
orders in $\theta$ expansion. 

Let us now come to the second possibility of getting the fermionic part of ${\bf B}^{(2)}_{mn}$ namely, S-dualizing ${\bf B}^{(1)}_{mn}$ i.e the NS part of 
the two-form (with its fermionic completion). In light of our earlier discussion,
this would be like moving up the type IIB coupling, at fixed self-dual radii of the compact spaces, so as to reach $g_s \to 1^-$ point. In other words, 
we are moving from region $B$ to region $A$ in {\bf figure \ref{IIBmod}}. 
 
We will however start by first fulfilling the promise that we made earlier, namely
discuss the appearance of $\sigma_3$, the third Pauli matrix, in \eqref{NorM} for 
the NS B-field ${\bf B}^{(1)}_{mn}$. Recall that 
our argument was to motivate the result from T-dualizing the metric component with cross-terms from type IIA to type 
IIB theory. Under T-duality the 32 component type IIA chiral fermion $\theta_A$ transforms as:
\bg\label{32comp}
&&\theta_A = \left(\begin{matrix} \theta_+ \\\theta_-\end{matrix}\right) ~ \to ~ 
\left(\begin{matrix} 1 & 0 \\0 & ~~~- \Gamma^{10} \Gamma_x \end{matrix}\right) \left(\begin{matrix} \theta_1 \\\theta_2 \end{matrix}\right) 
\equiv \widetilde{\Sigma}_1 \theta \nonumber\\
&&\bar{\theta}_A = \left(\begin{matrix} \bar{\theta}_+ & \bar{\theta}_-\end{matrix}\right) 
~ \to ~ \left(\begin{matrix} \bar{\theta}_1 & \bar{\theta}_2 \end{matrix}\right) 
 \left(\begin{matrix} 1 & 0 \\0 & ~~~~~ \Gamma^{10} \Gamma_x \end{matrix}\right) 
\equiv \bar{\theta} \widetilde{\Sigma}_2, \nd 
where the T-duality is performed along direction $x$ to go from IIA to IIB. The above transformations immediately implies the following algebra, similar to the
algebras that we discussed earlier in \eqref{algebras}:
\bg\label{2algeb} 
\widetilde{\Sigma}_2 \otimes \left(\begin{matrix} {\bf C}_{16 \times 16} & {} & 0\\ 0 & {} & {\bf C}_{16 \times 16} \end{matrix}\right) 
\otimes \widetilde{\Sigma}_1  ~ = ~ \left(\begin{matrix} {\bf C}_{16\times 16} & {} & 0 \\ 0 & {} & ~~~ - \Gamma^{10}\Gamma_x {\bf C}_{16 \times 16} 
\Gamma^{10} \Gamma_x\end{matrix}\right) = \sigma_3^p\otimes {\bf C}_{16\times 16}, \nonumber\\ \nd
where $\sigma_3$ is the third Pauli matrix with $p = 1$ or 2 depending on the specific 
representation of the 16-dimensional ${\bf C}$ matrix.
To fix the value of $p$, we can go to our self-dual point such that the transformation \eqref{32comp} becomes an intermediate transformation at $R_x = R_\perp = 1$,
where $R_\perp$ is the radius of an orthogonal circle.
We can choose ${\bf C}$ matrix to be of the form: ${\bf C}_{xm} ~ \equiv ~ \Gamma_x {\cal O}_m$, with ${\cal O}_m$ being a combination of type IIB fields and covariant derivatives with {\it even} or {\it odd} number of Gamma matrices. In that case $p =1$ in \eqref{2algeb}. 
Even
when the intermediate matrix, in the $\theta$ expansion, is of the form ${\bf C} + \sigma_1 \otimes \widetilde{\bf C}$, result of the form \eqref{2algeb} 
will continue to hold because we can absorb $\sigma_1$ in the transformation matrices as in \eqref{wayout}. Therefore, 
combining the results together, and assuming $p = 1$,
we can express the fermionic part of the NS B-field ${\bf B}_{mn}^{(1f)}$ as:
\bg\label{bnsas} 
{\bf B}_{mn}^{(1f)} =  \bar{\theta} \sigma_3 \otimes \Gamma^{10} M_{[mn]} \theta. \nd
As discussed earlier, we can now go to a corner of type IIB moduli space where the string coupling is strong i.e $g_s \to 1$. 
Here we expect the RR B-field ${\bf B}^{(2)}_{mn}$
to be given at least by the S-dual of ${\bf B}^{(1)}_{mn}$. The S-duality matrix that concerns us here is:
\bg\label{sdual}
\left(\begin{matrix} 0 & {} & -1\\ 1 & {} & ~~0\end{matrix}\right), \nd
which squares to $-{\bf I}_2$. This is the perturbative piece of the duality that keeps the string coupling unchanged, but changes the signs of the two-form fields. 
To incorporate S-duality in our fermionic part of the NS B-field ${\bf B}^{(1)}_{mn}$ one needs only to insert 
$-i\sigma_2$ in \eqref{bnsas} to give us the following fermionic piece\footnote{The sign is chosen for later convenience.}:
\bg\label{ferbrr}
\hat{\bf D}^{(2)}_{mn} = -i\bar{\theta} \sigma_3\sigma_2 \otimes \Gamma^{10} M_{[mn]} \theta, \nd
such that S-dualizing twice will yield $(-i\sigma_2)^2 = -{\bf I}_2$. This way we will get back the same result as \eqref{sdual} after two S-dualities that allow for
a ${\bf Z}_2$ phase factor.  
Combining \eqref{chat} and \eqref{ferbrr} together we get our final expression for the RR two-form
field along with its fermionic completion as:
\bg\label{rdrp}
{\bf B}^{(2)}_{mn} = C^{(2)}_{mn} -i\bar{\theta} \sigma_3 \sigma_2 \otimes \Gamma^{10}\left(M_{[mn]} 
+ i\Gamma_{mn} \widetilde{\Delta}^{(2)}\right)\theta. \nd
{}From the above expression we expect the fermionic terms to be suppressed by powers of string coupling {\it away} from the 
self-dual points, so that at the self-dual point (the region $A$ in {\bf figure \ref{IIBmod}}) we can 
exchange ${\bf B}^{(2)}$ and ${\bf B}^{(1)}$ and simultaneously perform two T-dualities. 
To see whether this is indeed true, we need to expand \eqref{rdrp} to higher orders in $\theta$. This can be
easily worked out using earlier expressions for $M_{[mn]}$ and $\widetilde{\Delta}$ in \eqref{mmn} and \eqref{rrfield} respectively, and the result is 
given by:
\bg\label{dhongsho} 
{\bf B}^{(2)}_{mn} &=&  C^{(2)}_{mn} -i \bar{\theta} e^{-\phi}\sigma_3 \sigma_2 \otimes \Gamma^{10}\left(M^{(11)}_{[mn]}  
+ i\Gamma_{mn} \widetilde{\Delta}^{(112)}\right)\theta \\
&-& i\bar{\theta} e^{-\phi}\sigma_3 \sigma_2 \otimes \Gamma^{10} \left(M^{(21)}_{[m\vert p} \theta \bar{\theta} M^{(22)}_{q\vert n]} g^{pq} + 
i\Gamma_{mn} \widetilde{\Delta}^{(212)}_{rp} \theta \bar{\theta} \widetilde{\Delta}^{(222)}_{qs} g^{pq} g^{rs}\right)\theta + {\cal O}(\theta^8), \nonumber \nd 
where to ${\cal O}(\theta^2)$ the coefficients can be read off from \eqref{1terma} and \eqref{jhor} as (see also \cite{marolf1, marolf2} for more details):
\bg\label{value}
M^{(11)}_{[mn]}  =  - \Gamma_{[m}\bar{\delta} \psi_{n]}, ~~~~~ \widetilde{\Delta}^{(112)} =  {1\over 2} \bar{\delta} \lambda. \nd 
We can see that the string coupling appears correctly in \eqref{dhongsho}
as to allow for the right behavior of the form-fields in the full IIB moduli space. The
fermion variations ($\bar{\delta}\psi_m, \bar{\delta} \lambda$) are with respect to either the original type IIB variables or the T-dual type IIB variables in our 
transformation scheme. Note that once we know the functional form of $\hat{\Delta}^{(ab2)}_{mn}$ for generic values of ($a, b$), we will know the 
$\theta$ expansion of \eqref{dhongsho} to arbitrary orders. This is of course a challenging exercise which we will not perform here. Instead we will use
our results for ${\bf B}^{(1)}_{mn}$ and ${\bf B}^{(2)}_{mn}$ etc to determine the fermionic structure of the four-form ${\bf C}_{mnpq}$ around the self-dual point. 

The fermionic structure of the four-form can be determined using similar trick as before by scanning the IIB moduli space. There are two differents points 
in the moduli space that would give us the four-form. First, at weak string coupling, we can go to the small compactification radii (or more appropriately the 
self-dual radii)
where the four-form can get 
contributions from the T-dual of ${\bf B}^{(2)}_{mn}$. Secondly, at strong string coupling i.e $g_s \to 1$, 
we can again go to self-dual radii where the  
four-form can now get contributions from the U-dual of ${\bf B}^{(1)}_{mn}$. For the first case, we can T-dualize twice the RR field ${\bf B}^{(2)}_{mn}$ 
along directions ($a, b$); and for the second case, we can S-dualize the ${\bf B}^{(1)}_{mn}$ field and then T-dualize twice along directions ($a, b$).  
The Gamma matrix algebra useful for us are now the following:
\bg\label{adore}
&& \left(\begin{matrix} 1 & {} & 0 \\ 0 & {} & ~~~~\Gamma_b \Gamma_a \end{matrix} \right) (\sigma_3 \sigma_2 \otimes \Gamma^{10}) \left(\begin{matrix} 1 & {} & 0 \\ 0 & {} & ~~~~\Gamma_a \Gamma_b \end{matrix} \right) = \sigma_2 \otimes \Gamma^{10} \Gamma_a \Gamma_b \\
&& \left(\begin{matrix} 1 & {} & 0 \\ 0 & {} & ~~~~\Gamma_b \Gamma_a \end{matrix} \right)( \sigma_3 \sigma_2 \otimes \Gamma^{10} \Gamma_x \Gamma_y )\left(\begin{matrix} 1 & {} & 0 \\ 0 & {} & ~~~~\Gamma_a \Gamma_b \end{matrix} \right) 
= \sigma_2 \otimes \Gamma^{10} \Gamma_a \Gamma_b \Gamma_x \Gamma_y. \nonumber \nd
Using these algebras, which are basically the T-duality rules, for both strong and weak string couplings will immediately provide us the contributions to the 
four-form from the two sources mentioned above around $g_s = R_a = R_b = 1$. The result is:
\bg\label{4form}
{\bf C}_{mnpq} &= & C^{(4)}_{mnpq} -i \bar{\theta} \sigma_2 \otimes \Gamma^{10} \left(2 \Gamma_{[mn} M_{pq]} 
+ i\Gamma_{mnpq} \widetilde{\Delta}^{(2)}\right)\theta \nonumber\\
&=&  C^{(4)}_{mnpq} -i\bar{\theta} \sigma_2 \otimes \Gamma^{10}\left(2 \Gamma_{[mn}M^{(11)}_{pq]}  
+ i\Gamma_{mnpq} \widetilde{\Delta}^{(112)}\right)\theta + {\cal O}(\theta^4)\nonumber\\
& = & C^{(4)}_{mnpq} - {1\over 2} \bar{\theta} \sigma_2 \otimes \Gamma^{10}\left(4\Gamma_{[mnp}\bar{\delta}\psi_{q]} - \Gamma_{mnpq}\bar{\delta}\lambda\right)\theta
+ {\cal O}(\theta^4),
\nd
where the factor of 2 signifies the contributions from the U-dual of the two B-fields, and we have determined the results upto ${\cal O}(\theta^2)$. One may 
verify with \cite{marolf1, marolf2, bergkallosh, shiu} that the result quoted above matches well with the literature at the self-dual point. 
It is interesting that to this order the match is
exact, therefore other possible corners of the type IIB moduli space do not contribute anything else to the fermionic parts of the bososnic RR and NS fields. 
At higher orders in $\theta$ there could be contributions that we cannot determine using out U-duality trick. Nevertheless, the U-duality transformations are
powerful enough to extract out the fermionic contributions from various corners of the moduli space.  

So far however we have not discussed the connection between $\Delta^{(1)}$ appearing in the dilaton and $\widetilde{\Delta}^{(2)}$ appearing in the axion, 
as in \eqref{oneroof}. The fact that they are related can be seen from M-theory on a torus ${\bf T}^2$ in the limit when the torus size is shrunk to zero.
Of course the scenario that we have envisioned here at the self-dual point cannot be uplifted to M-theory because we are not allowed to shrink the M-theory torus 
to zero size (as $g_s = 1$). However away from the self-dual point we {\it can} lift our configuration to M-theory, so let us discuss this point briefly.    
In M-theory we expect the metric to take a form similar to \eqref{meturi} or \eqref{oneroof}, i.e:
\bg\label{mmetric}
\hat{\bf G}_{mn} = G^{(11)}_{mn} + \bar{\theta} \hat{M}_{mn} \theta, \nd
where the superscript denotes the bosonic part of the metric, and $\theta$ is the corresponding fermionic variable. If we parametrize the torus direction by 
($x^3, x^a$) where $x^a$ denotes the eleventh-direction, then it is easy to see that in the limit of vanishing size of the torus, the type IIB axio-dilaton, with their
fermionic completions, are related via:
\bg\label{juddho}
{\rm exp}\left[-2{\bf \Phi}^{(1)}\right] + \left[{\bf \Phi}^{(2)}\right]^2 = {\hat{\bf G}_{33}\over \hat{\bf G}_{aa}}, \nd
implying the connection between $\Delta^{(1)}$ and $\widetilde{\Delta}^{(2)}$ {\it away from the self-dual point}. 
Using this one should be able to derive the ${\cal O}(\theta^2)$ result similar to
\eqref{dcmika} but away from the self-dual point,
as also given in \cite{marolf1, marolf2}. 
 
What happens at the self-dual point? The self-dual point is defined for $C^{(0)} = \phi = 0$, and therefore we should at least assume that this continues to be the 
case for the fermionic completions of the dilaton and axion too. In other words we should expect:
\bg\label{dilaxe}
{\bf \tau} \equiv {\bf \Phi}^{(2)} + i e^{-{\bf \Phi}^{(1)}}  =  i \;\;\; \mbox{(at the self-dual point),} \nd
to all orders in ($\theta, \bar{\theta}$).  Interestingly the condition $\vert {\bf \tau} \vert^2 = 1$ is similar to the M-theory condition \eqref{juddho} in the 
limit $\hat{\bf G}_{33} =  \hat{\bf G}_{aa}$. To lowest order in $\theta, \bar{\theta}$ it is easy to see that \eqref{dilaxe} reduces to the following condition:
\bg\label{jotadhar}
\bar{\theta}^\alpha\Delta^{(111)}_{\alpha\beta}\theta^\beta = - i \bar{\theta}^\alpha\left(\sigma_2\right)_{\alpha}^{\gamma} 
\widetilde{\Delta}^{(112)}_{\gamma\beta}\theta^\beta \;\;\; \mbox{(at the self-dual point).} \nd 
In general, to all orders in ($\theta, \bar{\theta}$), the relation between $\Delta^{(1)}$ and $\widetilde{\Delta}^{(2)}$ at the self-dual point 
can be directly seen from \eqref{dilaxe} as: 
\bg\label{dhour}
\bar{\theta} \Delta^{(1)}\theta = -{\rm log}\left(1+ i\bar{\theta} \sigma_2 \widetilde{\Delta}^{(2)}\theta\right)\;\;\; \mbox{(at the self-dual point)}. \nd
We expect \eqref{jotadhar} and \eqref{dhour} to 
reproduce the condition \eqref{known} or \eqref{dcmika} discussed in \cite{marolf1, marolf2, bergkallosh, martucci} at the self-dual point also. To this effect we will start by defining:
\bg\label{defdel}
\Delta^{(1)} = -i\widetilde{\Delta}^{(2)} + \hat{\Delta}, \nd
generically, both {\it at} and {\it away} from the self-dual point. Plugging \eqref{defdel} in \eqref{dhour}, and taking into account the lowest order results in 
\cite{marolf1, marolf2, bergkallosh, martucci}, we expect $\hat{\Delta}$ to vanish to lowest order in ($\bar{\theta}, \theta$) and the following constraint on the fermionic coordinate:
\bg\label{barts}
\bar{\theta}\left(1 - \sigma_2\right) = 0  \;\;\; \mbox{(at the self-dual point),}\nd
which would naturally explain the invariance under U-dualities at region $A$ in {\bf figure \ref{IIBmod}}. Of course away from the self-dual point we do not expect 
\eqref{dhour} and \eqref{barts} to hold, although \eqref{juddho} should continue to hold.

We now conclude this section by collecting together all of our results. The fermionic completions of the type IIB fields, away from the self-dual point, can 
be expressed in the following compact notations:
\bg\label{oneroof}
&& {\bf \Phi}^{(1)} = \phi + \bar{\theta} {\Delta}^{(1)} \theta, ~~~~{\bf \Phi}^{(2)} = C^{(0)} + \bar{\theta} e^{-\phi}\sigma_2 \widetilde{\Delta}^{(2)} \theta \nonumber\\
&& {\bf B}_{mn}^{(1)} = B_{mn} + \bar{\theta} \sigma_3\otimes \Gamma^{10}{M}_{[mn]} \theta,~~{\bf G}_{mn} = g_{mn} + \bar{\theta} M_{(mn)} \theta \nonumber\\ 
&& {\bf B}^{(2)}_{mn} = C^{(2)}_{mn} -i\bar{\theta} e^{-\phi}\sigma_3 \sigma_2 \otimes \Gamma^{10}\left(M_{[mn]} + i\Gamma_{mn} \widetilde{\Delta}^{(2)}\right)\theta \nonumber\\
&&  {\bf C}_{mnpq} =  C^{(4)}_{mnpq} -i\bar{\theta} e^{-\phi}\sigma_2 \otimes \Gamma^{10} \left(2\Gamma_{[mn} M_{pq]} 
+ i\Gamma_{mnpq} \widetilde{\Delta}^{(2)}\right)\theta, \nd
where the $\theta$ expansion for $\Delta^{(1)}$ is given by \eqref{bosfields}, for $\widetilde{\Delta}^{(2)}$ is given by \eqref{rrfield} and for 
$M_{(mn)}$ and $M_{[mn]}$ are given by \eqref{meturi}. We will take ($C^{(0)}, \phi$) $\to 0$, such that $g_s = e^\phi \to 1$ at the self-dual point.
Knowing these series expansions we can in principle determine the type IIB fields to arbitrary orders in
$\theta$ (provided of course there are no additional terms other than the ones got via U-duality transformations). In the presence of an ${\overline{\rm D3}}$, the functional forms for 
$\Delta^{(1)}, \widetilde{\Delta}^{(2)}$ and $M_{mn}$ become fixed. Henceforth this is the choice that we will consider, unless mentioned 
otherwise\footnote{For simplicity we will only concentrate 
on the integer $\overline{\rm D3}$ brane (including D3-brane), and not discuss the fractional branes as we did for the resolved conifold case. Although with our formalism it is easy to extend 
to any D-brane, integer or fractional, one needs to be careful when fractional branes are present along with integer D3 or $\overline{\rm D3}$-branes. 
However in the presence of only fractional branes, but no integer branes, the story proceeds in exactly the same way 
as discussed here as long as we are below the energy scale proportional to the inverse size of the two sphere on which we have our wrapped branes.}. 
For example, 
to ${\cal O}(\theta^2)$, $\Delta^{(1)}, 
\widetilde{\Delta}^{(2)}$ and $M_{mn}$ are known to be:
\bg\label{dcmika}
\Delta^{(1)} = -{i\over 2} \bar{\delta}\psi, ~~~~~ \widetilde{\Delta}^{(2)} = {1\over 2}\bar{\delta} \psi, ~~~~~ M_{mn} = -i \Gamma_m \bar{\delta}\psi_n, \nd
and therefore plugging them in \eqref{oneroof} will determine the type IIB fields to ${\cal O}(\theta^2)$ in the presence of an ${\overline{\rm D3}}$-brane. 
The above values should be understood as operators 
acting on $\theta$, and therefore to higher orders in $\theta$ one would need to express in terms of components:
\bg\label{compol} 
\left(\Delta^{(111)}_{\alpha\beta}, \Delta^{(ab1)}_{mn\alpha\beta}\right), ~~~~
\left(\widetilde{\Delta}^{(112)}_{\alpha\beta}, \widetilde{\Delta}^{(ab2)}_{mn\alpha\beta}\right), ~~~~ M^{(ab)}_{mn\alpha\beta}, \nd
as elucidated in \eqref{bosfields}, \eqref{rrfield} and \eqref{meturi} to properly write the higher order terms. Also, in \eqref{compol}
($m, n$) are Lorentz indices, and ($\alpha, \beta$) are spinor indices.
One may easily check that these results match with 
the ones known in the literature \cite{grana, trivedi, marolf1, marolf2, bergkallosh, martucci} for $e^\phi = 1$. 
The interesting thing about \eqref{compol} is that, knowing these
coefficients, one might be able to go to higher orders in $\theta$ as discussed above.

\subsection{$\kappa$-symmetry at all orders in $\theta$ }

In the previous section we managed to get the full fermionic action for the $\overline{\rm D3}$ branes using certain U-duality transformations at the self-dual
point in the type IIB moduli space. The result is extendable to the ${\rm D3}$-brane also, modulo certain subtleties that we want to elaborate here. 
Our answer is given in \eqref{fermc} which is derived for the special case of ${\cal F}_{mn} = 0$.  The most generic case,
given as \eqref{d3ad3}, could also be worked out using the representations \eqref{oneroof} for the type IIB
fields, but we will not do so here. 

Another issue that we briefly talked about earlier is the behavior of these 
higher order terms under Renormalization Group flow. Under RG we expect these terms to be
irrelevant. However as we will discuss momentarily, to argue for the full $\kappa$-symmetry, all the higher order terms are essential. Therefore for our purpose it may be useful to
work with the {\it exact} renormalization group equations \cite{polrez} to keep track of the irrelevant operators. In the following however we will not discuss the quantum behavior 
and concentrate only on the classical action \eqref{fermc} with all the higher order terms. 

The question that we want to answer here is the following: under what condition will the action \eqref{fermc} take the $\kappa$-symmetric form, i.e a form 
like $\mathcal{L} \sim \bar{\theta}(1 - {\bf \Gamma}^{\pm}_{\rm D3})[\,...\,]\theta$, where ${\bf \Gamma}^{\pm}_{\rm D3}$ is the $\kappa$-symmetry 
operator\footnote{See \eqref{gD3} for the definition of ${\bf \Gamma}^{\pm}_{\rm D3}$.}? The condition, as we shall see, turns out to be rather subtle so we will have to tread carefully. 
Therefore as a start we will take the world-volume action, for a single D3 or $\overline{\rm D3}$, in the presence of the fermionic terms, to be given by:
\bg\label{d3ad3}
S = -T_3\int d^4\zeta e^{-{\bf \Phi}^{(1)}}\sqrt{-{\rm det}\left({\bf G}_{ab} + {\bf B}^{(1)}_{ab} + \alpha'{\bf F}_{ab}\right)} \pm T_3 \int {\bf C} \wedge 
e^{{\bf B} + \alpha'{\bf F}}, \nd 
where the first term is the Born-Infeld (BI) piece and the second one is the Chern-Simons (CS) piece. The only difference now is that they both include the 
fermionic completions that we developed earlier which are in general different for D3 and $\overline{\rm D3}$ branes\footnote{We have used three kind of matrices, 
namely $M_{mn}, \Delta^{(1)}$ and $\widetilde{\Delta}^{(2)}$ to 
express the fermionic pieces in the presence of an $\overline{\rm D3}$-brane. One may choose similar matrices to express the fermionic pieces in the presence of a 
D3-brane. For example we will use $M^+_{mn}, \Delta^{(1+)}$ and $\widetilde{\Delta}^{(2+)}$ 
as the corresponding matrices for a D3-brane to represent the 
fermionic parts, whereas $M^-_{mn} = M_{mn}, \Delta^{(1-)}= \Delta^{(1)}$ and $\widetilde{\Delta}^{(2-)} = \widetilde{\Delta}^{(2)}$ will be reserved 
for the $\overline{\rm D3}$-brane to avoid clutter. \label{context}}. 
We can choose the gauge field ${\bf F}_{ab}$ in such a way as to cancel the fermionic contributions of the 
NS B-field ${\bf B}_{ab}^{(1)}$. This way we can write a bosonic combination ${\cal F}_{ab} \equiv {\bf B}^{(1)}_{ab} + \alpha'{\bf F}_{ab}$ to represent the 
gauge field. We can also define a matrix $A$ in the following way:
\bg\label{adefinu}
A_{mn} \equiv \left[\left(g + {\cal F}\right)^{-1}\right]^p_m \bar{\theta}^\alpha M_{pn\alpha\beta}\theta^\beta, \nd
with $M_{mn}$ matrix defined earlier in \eqref{oneroof} to study the fermionic parts of the metric and the NS B-field. With this definition, the BI part of the 
antibrane action takes the following form:
\bg\label{bi} 
S_{\rm BI} = -T_3 \int d^4\zeta e^{-\phi}\sqrt{-{\rm det}(g + {\cal F})} ~{\rm exp}\left[{1\over 2} {\rm tr}~{\rm log}\left({\bf I} + A\right) - \bar{\theta} 
\Delta^{(1)}\theta\right], \nd
where ${\bf I}$ is the identity matrix in four-dimension, and $A$ is the same matrix defined earlier in \eqref{adefinu}. As usual, at the self-dual point we put $\phi =0$ to be 
consistent with our U-dualities. Moving away from the self-dual points, as exemplified in \eqref{dhongsho}, \eqref{value} and \eqref{oneroof}, the action has the necessary
dilaton piece. 

We now come to the Chern-Simons part of the brane action for both the D3 and $\overline{\rm D3}$ using the fermionic completions developed above. The action can be written as:
\bg\label{csa}
S_{\rm CS} = T_3 \int d^4\zeta \epsilon^{mnpq} \left({\bf C}^{\pm}_{mnpq} + {\bf B}^{(2\pm)}_{mn} {\cal F}_{pq} 
+ {1\over 2} {\bf \Phi}^{(2\pm)} {\cal F}_{mn} {\cal F}_{pq}\right), \nd
where the superscript represent D3 and $\overline{\rm D3}$ respectively, and 
${\bf C}^{-}_{mnpq} \equiv {\bf C}_{mnpq}$, ${\bf B}^{(2-)}_{mn} \equiv {\bf B}^{(2)}_{mn}$ and ${\bf \Phi}^{(2-)} \equiv {\bf \Phi}^{(2)}$ for an 
$\overline{\rm D3}$ as we developed here.
We have assumed that the background is flat along spacetime directions so that the curvature terms do not appear above. In general, for curved background, the
curvature terms with their fermionic completions (from the metric) should also appear. For our case this should only change the last term in the above action \eqref{csa}.

We can simplify the action \eqref{csa} further by assuming ${\cal F}_{mn} = 0$. This would also imply that $A_{mn}$ in \eqref{adefinu} simplifies. This is the case we will 
consider here. A more generic scenario with ${\cal F}_{mn}$, or even with the fermionic pieces of ${\cal F}_{mn}$ (that we cancelled here) can be studied. 
This will make the system more involved but won't change the physics. Therefore, for this special case we have:
\bg\label{jmaro}
S_{\rm CS} = T_3 \int d^4\zeta \epsilon^{mnpq} C^{(4)}_{mnpq} 
+ T_3 \int d^4\zeta e^{-\phi}\sqrt{-{\rm det}~g} ~\bar{\theta}~ {\bf \Gamma}^{\pm}_{\rm D3}\left({1\over 2} {\Gamma}^{ba} M^{\pm}_{ab} 
+ i\widetilde{\Delta}^{(2\pm)}\right)\theta, \nd
where, as before, $M^{-}_{ab} \equiv M_{ab}$ and $\widetilde{\Delta}^{(2-)} \equiv \widetilde{\Delta}^{(2)}$ represent the 
corresponding matrices for an $\overline{\rm D3}$; and ${\bf \Gamma}^{\pm}_{\rm D3}$ is defined as:
\bg\label{gD3}
{\bf \Gamma}^{\pm}_{\rm D3} = \pm {i\sigma_2 \otimes \Gamma^{10} \Gamma_{mnpq}\epsilon^{mnpq}\over 4!\sqrt{-{\rm det}~g}}. \nd
Let us now come back to the BI piece of the action \eqref{bi}. To analyze this we will use the well known expansion for log as:
\bg\label{log}
{\rm tr} ~{\rm log}\left({\bf I} + A\right) = {\rm tr}~A - {1\over 2} {\rm tr}~A^2 + {1\over 3} {\rm tr}~A^3 + .... 
= \sum_{k=1}^{k_{max}} {(-1)^{k+1} {\rm tr}~ A^k\over k}, \nd
where $k_{max}$ is determined by the rank of the matrix. Plugging this in the BI action \eqref{bi} and rearranging the action appropriately, we get for an 
$\overline{\rm D3}$:
\bg\label{binow}
S_{\rm BI} & = &  -T_3 \int d^4\zeta e^{-\phi}\sqrt{-{\rm det}~g} \left[1 + \sum_{k = 1}^{k_{max}}\left({1\over 2}{\rm tr}~A - \bar{\theta}\Delta^{(1)}\theta 
-{1\over 2} \sum_{l = 1}^{l_{max}}{{\rm tr}~(-A)^{l+1}\over l}\right)^k\cdot {1\over k!}\right]\nonumber\\
& = & - T_3 \int d^4\zeta e^{-\phi}\sqrt{-{\rm det}~g}~ \left[1 + \sum_{k= 1}^{k_{max}} {\left({1\over 2} {\rm tr}~A + i\bar{\theta} \widetilde{\Delta}^{(2)}
\theta + {\cal S}(A, \hat{\Delta})\right)^k\over k!}\right] 
\nd 
where the first term is the standard BI term for the bosonic piece and the second term is the fermionic extension. 
We have also used \eqref{defdel} to replace $\Delta^{(1)}$ by $\widetilde{\Delta}^{(2)}$ and defined the other variable appearing above in the following way:
\bg\label{vardef}
{\cal S}(A, \hat{\Delta}) = -{1\over 2} \sum_{l = 1}^{l_{max}}{{\rm tr}~(-A)^{l+1}\over l} - 
\bar{\theta} \hat{\Delta} \theta. \nd
Combining the Chern-Simons and the Born-Infeld parts, i.e \eqref{jmaro} and \eqref{binow} respectively, we can extract the fermionic completions of the 
brane and anti-brane actions. The result is given by: 
\bg\label{fermc}
&&S^f_{\pm} = - T_3 \int d^4\zeta e^{-\phi}\sqrt{-{\rm det}~g}~ {\cal L}_{\pm}\\
&& {\cal L}_{\pm} \equiv \left[\sum_{k= 1}^{k_{max}} {\left({1\over 2} {\rm tr}~A^{\pm} + i\bar{\theta} \widetilde{\Delta}^{(2\pm)}
\theta + {\cal S}^\pm(A, \hat{\Delta})\right)^k\over k!} -
\bar{\theta}~ {\bf \Gamma}^\pm_{\rm D3}\left({1\over 2} {\Gamma}^{ba} M^\pm_{ab} + i\widetilde{\Delta}^{(2\pm)}\right)\theta \right], \nonumber \nd
where $\pm$ subscript denote D3 brane and $\overline{\rm D3}$ respectively and $A^-_{mn} \equiv A_{mn}$ as in \eqref{adefinu}. 
The bosonic parts of the action for the brane and the anti-brane remain the same as 
the standard ones, as one can easily verify. It is also easy to see that:
\bg\label{tA}
{1\over 2}{\rm tr}~A^\pm = {1\over 2} \bar{\theta} \Gamma^{ba}M^\pm_{ab}\theta \equiv \bar{\theta} \left({\bf N}_\pm - i\widetilde{\Delta}^{(2\pm)}\right)\theta, \nd
where ${\bf N}_\pm$ is defined in such a way that the fermionic action \eqref{fermc} takes the following form:
\bg\label{spm}
S_{\pm}^f = -T_3 \int d^4\zeta e^{-\phi}\sqrt{-{\rm det}~g} \left(e^{\bar{\theta}{\bf N}_\pm\theta + {\cal O}({\bf N}_\pm^2)}-1 -\bar{\theta} {\bf \Gamma}^\pm_{\rm D3} {\bf N}_\pm
\theta\right). \nd  
In the absence of any other information about the series ${\bf N}_\pm$, the above action for the fermionic terms for the D3 or the $\overline{\rm D3}$ is probably the
best we can say at this stage. 
Simplification can occur when ${\bf N}_\pm$ remains small to 
all orders in ($\theta, \bar{\theta}$), which in-turn would guarantee the smallness of the ${\cal O}({\bf N}_\pm^2)$ terms in the exponential, as well as 
the exponential itself. If this is the case then:
\bg\label{hotat}
S^f_\pm & = & -T_3 \int d^4\zeta e^{-\phi}\sqrt{-{\rm det}~g}~\bar{\theta} \left(1 +{\bf \Gamma}^\pm_{\rm D3}\right) {\bf N}_\pm \theta \nonumber\\
& = & -T_3 \int d^4\zeta e^{-\phi} \sqrt{-{\rm det}~g}~\bar{\theta} \left(1 + {\bf \Gamma}^\pm_{\rm D3}\right) 
\left({1\over 2} {\Gamma}^{ba} M^\pm_{ab} + i\widetilde{\Delta}^{(2\pm)}\right)\theta, \nd 
which would provide a strong confirmation of the recent work of \cite{wrase1}, which was originally done to ${\cal O}(\theta^2)$. For our case we can use the 
$\theta$-expansions for $M^-_{ab} = M_{ab}$ and $\widetilde{\Delta}^{(2-)} = \widetilde{\Delta}^{(2)}$ for an $\overline{\rm D3}$ to express:
\bg\label{chappal}
\bar{\theta} \left({1\over 2} {\Gamma}^{ba} M_{ab} + i\widetilde{\Delta}^{(2)}\right)\theta  &=&  
\bar{\theta}^\alpha \left({1\over 2} {\Gamma}_{\alpha}^{ba\gamma} M^{(11)}_{ab\gamma\beta} + i\widetilde{\Delta}^{(112)}_{\alpha\beta}\right)\theta^\beta \nonumber\\
& + & \bar{\theta}^\alpha\left({1\over 2} \Gamma^{ba\gamma}_{\alpha}M^{(21)}_{ac\gamma\delta}\theta^\delta \bar{\theta}^\sigma M^{(22)c}_{b\sigma\beta} + 
i\widetilde{\Delta}^{(212)}_{\alpha\delta m}\theta^\delta \bar{\theta}^\sigma \widetilde{\Delta}^{(222)m}_{\sigma\beta}\right)\theta^\beta + {\cal O}(\theta^6)
\nonumber\\
&=& -{1\over 2} i\bar{\theta} \left(\Gamma^a \bar{\delta}\psi_a - \bar{\delta}\lambda\right) \theta + {\cal O}(\theta^4), \nd
which is consistent with what we know to ${\cal O}(\theta^2)$ from the literature \cite{marolf1, marolf2, bergkallosh, shiu}. Now if we define 
${\bf \Gamma}^-_{\rm D3} = {\bf \Gamma}_{\rm D3}$ and ${\bf \Gamma}^+_{\rm D3} = -{\bf \Gamma}_{\rm D3}$ from \eqref{gD3} and 
$\delta^+ = \delta$ and $\delta^- = \bar{\delta}$ from \cite{wrase1};
and using 
the fermionic actions \eqref{fermc} or \eqref{hotat} for the D3 and the $\overline{\rm D3}$ branes, then to ${\cal O}(\theta^2)$ we can easily 
reproduce the expected result in $\kappa$-symmetric form: 
\bg\label{d3aad3}
S_{\pm} = {1\over 2} T_3 \int d^4\zeta e^{-\phi} \sqrt{-{\rm det}~g} ~i\bar{\theta} \left(1 \mp {\bf \Gamma}_{\rm D3}\right) 
\left(\Gamma^a \delta^\pm\psi_a - \delta^\pm \lambda\right)
\theta + {\cal O}(\theta^4). \nd
At the orientifold point, if we assume that the action is given by \eqref{hotat}, then to all orders in $\theta$ the fermionic coordinate satisfy 
$\bar{\theta}\left(1 - {\bf \Gamma}_{\rm D3}\right) = 0$. This way $S_+$ vanishes identically and $S_-$ remains non-zero. This result seems to be valid only if the 
fermionic action takes the form \eqref{hotat}, but is not obvious from the fermionic action \eqref{fermc} that this will continue to be the case. 
In fact the action \eqref{fermc} has many terms, coming from
the log and from the exponential pieces, that do not in any obvious way give us $S_+ = 0$ at the orientifold point. In the following we will try to see how 
we can adjust the background, for example \eqref{oneroof}, to get the required form of the action.

Clearly adjusting the background should effect the definition of the type IIB fields \eqref{oneroof}. From the way we derived \eqref{oneroof}, we cannot arbitarily change the field definitions since they are related by certain U-duality transformations at a self-dual point. Thus for example, knowing ${\bf B}^{(1)}_{mn}, {\bf \Phi}^{(1)}$ and ${\bf \Phi}^{(2)}$, we pretty much derived the rest of the RR fields using U-dualitites. All the fields and their corresponding fermionic completions depend on three set of functional forms: $M_{mn}, \Delta^{(1)}$ and $\widetilde{\Delta}^{(2)}$. In fact the anti-symmetric part of the operator $M_{mn}$, namely $M_{[mn]}$, is essential to describe the fermionic completions of the $p$-form fields in type IIB. The symmetric part, $M_{(mn)}$, on the other hand is reserved for the fermionic completion of the metric. At the self-dual radii, $M_{(mn)}$ and $M_{[mn]}$, could be related by T-dualities along one parallel and one orthogonal spatial directions. The temporal directions however are not connected via simple T-dualities. This distinction may help us to construct the $\kappa$-symmetric form of the action from \eqref{fermc}. To this end, we start by
redefining the temporal components of the metric ${\bf G}_{0\mu}$ in the following way:
\bg\label{tempco}
{\bf G}_{00} \equiv \left(g_{00} + \bar{\theta} M_{00}\theta\right) {\rm exp}\left(2\bar{\theta}\Omega\theta\right), ~~~
{\bf G}_{0i} \equiv \left(g_{0i} + \bar{\theta} M_{0i}\theta\right){\rm exp}\left({1\over 5}\bar{\theta}\Omega\theta\right), \nd
keeping ${\bf G}_{ij}$ and all other type IIB fields exactly as in \eqref{oneroof}. The $\Omega(\theta, \bar{\theta})$ appearing above is again a series defined by
powers of ($\theta, \bar{\theta}$) as:
\bg\label{omg}
\bar{\theta}\Omega\theta  = \bar{\theta}^\alpha \Omega^{(11)}_{\alpha\beta}\theta^\beta + 
\bar{\theta}^\alpha \Omega^{(21)}_{m...q\alpha\gamma}\theta^\gamma   \bar{\theta}^\delta \Omega^{(22)}_{p...n\delta\beta}\theta^\beta g^{qp}...g^{mn} + {\cal O}(\theta^6)\nd   
where the coefficients can be defined in a similar way as the variables appearing in \eqref{oneroof}. As before, we could resort to rank two tensor representations for $\Omega^{(21)}$
and $\Omega^{(22)}$ etc., without losing much of the physics here. 

Let us now revisit the Born-Infeld part of the action \eqref{d3ad3}. Taking \eqref{tempco} and \eqref{oneroof} into account, it is easy to see that the BI action now takes
the following form:
\bg\label{jutabez}
S_{BI} & = & -T_3\int d^4\zeta e^{-{\bf \Phi}^{(1)}}\sqrt{-{\rm det}\left({\bf G}_{ab} + {\bf B}^{(1)}_{ab} 
+ \alpha'{\bf F}_{ab}\right)}\bigg{\vert}_{{\bf B}^{(1)}_{ab} + \alpha'{\bf F}_{ab} \equiv 0} \\
& = & -T_3 \int d^4\zeta e^{-\phi} \sqrt{-{\rm det}~g}~{\rm exp}\left[{1\over 2}{\rm tr}~{\rm log}\left({\bf I} + A\right) + i \bar\theta \widetilde{\Delta}^{(2)} \theta
- \bar\theta \hat{\Delta} \theta + \bar\theta \Omega \theta\right] \nonumber\\
& = & -T_3\int d^4\zeta e^{-\phi} \sqrt{-{\rm det}~g}~{\rm exp}\left[{1\over 2} {\rm tr}~A + i \bar\theta \widetilde{\Delta}^{(2)}\theta - \left({1\over 2} \sum_{k = 2}^{k_{max}}
{(-1)^k~{\rm tr}~A^k\over k} + \bar\theta \hat\Delta \theta\right) + \bar\theta \Omega \theta\right]\nonumber\\
& \equiv & -T_3 \int d^4\zeta e^{-\phi}\sqrt{-{\rm det}~g}~{\rm exp}\left[\sum_{k = 1}^{k_{max}} {(-1)^{k+1}\over k}\left({1\over 2}{\rm tr}~A 
+ i\bar\theta \widetilde{\Delta}^{(2)}\theta\right)^k + \bar\theta \left(\Theta + \Omega\right)\theta\right] \nonumber 
\nd 
where going from the second-last to the last line of \eqref{jutabez}, we have used the following mathematical identity:
\bg\label{folide}
{1\over 2} \sum_{k = 2}^{k_{max}} {(-1)^k {\rm tr}~A^k\over k} + \bar\theta \hat\Delta \theta \equiv 
\sum_{k = 2}^{k_{max}} {(-1)^{k}\over k}\left({1\over 2}{\rm tr}~A 
+ i\bar\theta \widetilde{\Delta}^{(2)}\theta\right)^k + \bar\theta \Theta \theta, \nd
implying that the functional forms of $\Theta$  and $\hat\Delta$ can be used to express all ${\rm tr}~A^k$ in terms of $\left({\rm tr}~A\right)^k$ to allow for \eqref{folide}.
Additionally, since $\Omega$ in \eqref{tempco} is arbitrary, we can as well as absorb $\Theta$ in the definition of $\Omega$ to give us:
\bg\label{thom}
\bar\theta \left(\Theta + \Omega\right)\theta ~ = 0.\nd
The above two conditions \eqref{folide} and \eqref{thom} are essential for expressing the $\overline{\rm D3}$-brane action in the $\kappa$-symmetric form. Putting 
\eqref{folide} and \eqref{thom} in \eqref{jutabez}, we get:
\bg\label{binu}
S_{BI} &=& -T_3\int d^4\zeta e^{-{\bf \Phi}^{(1)}}\sqrt{-{\rm det}~{\bf G}_{ab}}\\   
& = & -T_3 \int d^4\zeta e^{-\phi} \sqrt{-{\rm det}~g} -T_3 \int d^4\zeta e^{-\phi} \sqrt{-{\rm det}~g} \, \bar\theta\left({1\over 2}\Gamma^{ba}M^\pm_{ab} 
+ i  \widetilde{\Delta}^{(2)}\right)\theta, \nonumber \nd
which is precisely the condition that is required for the BI action to take the $\kappa$-symmetric form when combined with the Chern-Simons part of the action \eqref{jmaro}. 
Thus putting \eqref{binu} and \eqref{jmaro} together, we get our final expression for the $\overline{\rm D3}$-brane action:
\bg\label{reenaroy}
S_{\overline{\rm D3}} = && -T_3 \int d^4\zeta e^{-\phi} \sqrt{-{\rm det}~g} - T_3 \int d^4\zeta ~\epsilon^{mnpq} C^{(4)}_{mnpq} \nonumber\\
&& -T_3 \int d^4\zeta e^{-\phi} \sqrt{-{\rm det}~g}~\bar{\theta} \left(1 - {\bf \Gamma}^-_{\rm D3}\right) 
\left({1\over 2}\Gamma^{ba}M^- _{ab} + i\widetilde{\Delta}^{(2)}\right)\theta, \nd 
in a manifestly $\kappa$-symmetric form. Equivalently, 
the above action indicates that the $\overline{\rm D3}$ $\kappa$-symmetry projector
\begin{equation}
\left(1 - {\bf \Gamma}^-_{\rm D3}\right) ,
\end{equation}
continues to be the $\kappa$-symmetry projector at $\emph{all}$ orders in $\theta$. Recall that the $\kappa$-symmetry variation of $\bar{\theta}$ is given by
\begin{equation}
\delta_\kappa \bar{\theta} = \bar{\kappa} (1 +{\bf \Gamma}^-_{\rm D3}  ).
\end{equation}
It follows from this that $\overline{\rm D3}$ action is manifestly $\kappa$-symmetric at \emph{all orders} in $\theta$. 

In deriving our result we have relied on the fact that at the self dual point we do not have extra fermionic operators other than the ones given by our U-duality transformations. This seems to
be the case in any given background, otherwise we will end up with extra fermionic condensates which would appear to violate equations of motion. On the other hand, the U-duality rules that we 
used here also have $\alpha'$ corrections \cite{bergkal, berg, olaf}
so one might worry that this could change our result. A careful thought will tell us that this is not the case, as in deriving our results we have
only used generic properties of T-duality. To see this in more details, let us investigate the two key relations where some aspects of the T-duality rules have been used, namely \eqref{NorM} 
and \eqref{rere}. The first relation i.e \eqref{NorM} relates $N^{(1)}_{[mx]}$ with $M_{[mx]}$ under one T-duality along direction $x$. This is one of the Buscher's rule derived for 
the limit $\alpha' \to 0$, so one would ask what happens under $\alpha'$ corrections. Before we go about discussing $\alpha'$ corrections to this, let us ask what does it mean to 
have a relation like \eqref{NorM}. Since the piece $M_{mn}$ comes from the metric and the piece $N_{mn}$ comes from the NS B-field, the relation, or at least the bosonic part of it, 
implies the connection between the momentum and the winding modes under one T-duality. Thus, this is in the spirit of charge conservation: momentum charges being exchanged with winding charges or 
vice-versa and we can take this to be the defining property of T-duality. Since \eqref{NorM} implies the fermionic version of this, we will assume that \eqref{NorM} do not have any 
additional $\alpha'$ pieces. 

Similar argument unfortunately
cannot be given for \eqref{rere}, where the RR two-form appears from the axion under two T-dualities, as unlike the previous argument $-$ 
where momentum and winding modes appear automatically $-$ we do not have the advantage of invoking charge conservation {\it a priori}. We do 
however notice that there is a possible {\it alternative}
way of expressing the fermionic parts of the background fields, namely that the background fields 
are functions of ($\theta, \bar{\theta}$) with the tensorial parts being specified by certain functions of the
spacetime coordinates. In this language the T-duality rules are simply given by the way ($\theta, \bar{\theta}$) change, i.e the transformation rules given in \eqref{thetachange}. This
way we don't have to worry about the explicit $\alpha'$ dependences appearing from the T-duality transformations, and the all-order result \eqref{oneroof} should be exact with the 
$\alpha'$ dependences now appearing from the order-by-order expansions of the ($\theta, \bar{\theta}$) terms for every components of the type IIB fields in \eqref{oneroof}.

\section{Conclusion and Discussion}

In this work we have studied the interplay of $\mathcal{N}=1$ supersymmetric backgrounds and anti-branes. We found two new examples where supersymmetry is spontaneously broken by a probe anti-brane: a $\overline{\rm D3}$ in a resolved conifold, and a $\overline{\rm D7}$ in a GKP background. In the first case, the low-energy spectrum in the probe approximation has two massless fermions. However, once backreaction of the $\overline{\rm D3}$ on bulk fluxes is taken into account (perturbatively), the would-be massless fermions in fact become massive; this is a consequence of having a wrapped five-brane in the background (an issue which does not arise when studying GKP-type backgrounds). In the second case, we found there can in fact be \emph{many} massless fermions, and the precise number depends on the Hodge numbers of the 4-cycle wrapped by the $\overline{\rm D7}$, although we did not extend the analysis to include backreaction. We also studied the effect of worldvolume fluxes, which provide extra mass terms. It is possible that for the most general worldvolume fluxes background there are no $\overline{\rm D7}$ fermions which remain massless. 

As a step towards a more complete understanding of anti-branes and supersymmetry breaking, we studied the brane fermionic action at all orders in the fermionic expansion. In other words, we studied the all-order $\alpha'$ expansion of the fermionic action, while working at \emph{leading order} in the bosonic $\alpha'$ expansion. This allowed us to neglect curvature corrections to the action, as well as purely bosonic $\alpha'$ corrections to the string duality transformations. Our result is that the all-order fermionic action can be written in a manifestly $\kappa$-symmetric form, which implies that our previous two analyses (and the results of \cite{wrase1,wrase2}) are not simply a leading-order effect. In this analysis we neglected the effect of worldvolume flux, and while we don't expect this to qualitatively change the result (see, for example, \cite {martucci}), it would be interesting to see the precise details of how this changes the all-order fermionic calculation.

There are many directions for future work. It would be interesting to see what types of inflationary scenarios can be built from the two examples we have studied, and if the interaction of the fermions with worldvolume fluxes can lead to a modification of the inflationary dynamics. In a totally different direction, we would like to see how the all-order fermionic action can be expressed in a Volkov-Akulov form, which should in principle be possible given the recent results of \cite{kuzen}. We plan to study all these effects in future works.

\vskip.2in

\noindent {\bf Acknowledgements}

\vskip.1in

\noindent It is our great pleasure to thank Eric Bergshoeff, Renata Kallosh and Antoine Van Proeyen for helpful discussions, and Timm Wrase for intial collaborations and many 
illuminating discussions. K. D would like to thank the Stanford Physics department for hospitality during his sabbatical visit, where this work was started. The work of 
K. D is supported in part by the National Science and Engineering Research Council of Canada and in part by the Simons Research Grant. The work of E. M is supported by the 
National Science and Engineering Research Council of Canada via a PGS D fellowship.

{}
 \end{document}